\documentclass[extra,referee]{gji}

\usepackage{timet}
\usepackage{url} 
\usepackage{lineno}
\usepackage{soul}
\usepackage{nccmath}
\usepackage{enumitem}
\usepackage{ifthen}
\usepackage{color}
\usepackage{graphicx}
\usepackage{amsmath}
\usepackage{amssymb}
\usepackage{algorithm}
\usepackage{algorithmic}
\usepackage{setspace}
\usepackage{multirow}
\usepackage[T1]{fontenc}

\graphicspath{{Fig/}}
\title[Designing Solutions to Geophysical Inverse Problems]{Designing Solutions to Geophysical Inverse Problems by Changing Variables}
\author[Zhao et al., 2026]
{Xuebin Zhao$^1$*, Andrew Curtis$^{1}$* and Klaus Mosegaard$^{2}$*\\
	$^1$ School of GeoSciences, University of Edinburgh, United Kingdom \\
	$^2$ Niels Bohr Institute, University of Copenhagen, Denmark \\
	$^*$ Equal contribution\\
	E-mail: \textit{xuebin.zhao@ed.ac.uk, andrew.curtis@ed.ac.uk}
}

\begin{document}
\label{firstpage}


\begin{titlepage}
	\begin{center}
		\vspace*{1.0cm} 
		
		\huge
		\textbf{Designing Solutions to Geophysical Inverse Problems by Changing Variables}
		
		\vspace{1.0cm}
		\LARGE
		Xuebin Zhao$^1$*, Andrew Curtis$^{1}$* and Klaus Mosegaard$^{2}$*\\
		
		\vspace{0.5cm}
		\Large
		$^1$ School of GeoSciences, University of Edinburgh, United Kingdom \\
		$^2$ Niels Bohr Institute, University of Copenhagen, Denmark \\
		$^*$ Equal contribution\\
		
		\vspace{0.4cm}
		\Large
		E-mail: \textit{xuebin.zhao@ed.ac.uk, andrew.curtis@ed.ac.uk}
		
		\vfill
		\vfill
	\end{center}
\end{titlepage}

\newpage

\begin{summary}
Geoscientists often solve inverse problems to estimate values of parameters of interest given relevant data sets. Bayesian inference solves these problems by combining probability distributions that describe uncertainties in both observations and unknown parameters, and we require that the solution provides unbiased uncertainty estimates in order to inform risk-based decisions. It has been known for over a century that employing different, but equivalent parametrisations of the same information can yield conditional probabilities that are mathematically inconsistent, a property referred to as the \textit{BK-inconsistency}. Recently this inconsistency was shown to invalidate the solutions to physical problems found using several well-established methods of Bayesian inference. In this study, we explore the extent to which this inconsistency affects solutions to common geophysical problems. We demonstrate that changes in parametrisations result in inconsistent conditional prior probability densities, even though they represent exactly the same prior information. We show that these inconsistent prior distributions can change Bayesian posterior solutions dramatically across various geoscientific problems including seismic impedance inversion, surface wave dispersion inversion, and travel time tomography, using real and synthetic data. Significantly different posterior statistics are obtained, including for maximum \textit{a posteriori} (MAP) solutions, mean estimates, standard deviations, and full posterior distributions. Given that deterministic inversion is often equivalent to finding the MAP solution to specific Bayesian problems (the mathematical equations to be solved are identical), the BK-inconsistency also results in inconsistent solutions to deterministic inverse problems. Indeed, we show that solutions can potentially be \textit{designed}, simply by changing the parametrisation. This study highlights that a careful rethinking of Bayesian inference and deterministic inversion may be required in physical problems: the effects that we demonstrate are likely to affect past and present inverse problem solutions in a variety of different fields of application.
\end{summary}

\section{Introduction}
Geoscientists usually estimate properties of the Earth system using observations of physical quantities. Transferring information in the observations to the target parameters of interest often involves solving an inverse or inference problem. Random noise in recorded data, the inherent nonlinearity of inverse problems, and practical constraints on the design of experiments or surveys creates significant non-uniqueness in the solutions. Accurate results and unbiased uncertainty assessments are therefore crucial if we are to interpret Earth properties and processes correctly, particularly if they are to be used to make informed, evidence or risk based decisions \citep{robert2007bayesian, berger2013statistical, arnold2018interrogation}.

Geoscientific inverse problems are commonly solved using methods in which the nonlinear forward function is linearised (approximated) iteratively around the best available model estimate. In principle, an optimal solution can then be derived using gradient-based optimisation algorithms \citep{boyd2004convex}. However, these methods usually require \textit{ad hoc} regularisation to stabilise the optimisation process and improve convergence rates \citep{zhdanov2002geophysical, asnaashari2013regularized}. In addition, a good initial model is required to prevent convergence to a locally optimal solution that is unrelated to the true Earth. Importantly, linearised methods can not provide reliable uncertainty estimates: while uncertainty information can be approximated by calculating the inverse Hessian matrix at any solution, this only captures local uncertainty near that solution and fails to address the full uncertainty created by the nonlinearity of most Geoscientific problems \citep{tarantola2005inverse}.

An alternative approach to solving inverse problems employs probabilistic, nonlinear methods using Bayes' rule, commonly known as \textit{Bayesian inference} or \textit{Bayesian inversion}. In this framework, both unknowns and observations are treated as random variables, and their uncertainties are characterised by probability distributions. During inversion, \textit{a priori} knowledge about the Earth is updated with information from the observed data to calculate the \textit{posterior} probability density function (pdf), a distribution that encapsulates all possible Earth models that fit the data to within their observational uncertainties. Uncertainties in the inversion results can thus be quantified, at least in principle \citep{tarantola2005inverse}.

In many cases, researchers reparametrise available information, either by using different model parameters, or different ways to describe or make observations. There can be a variety of reasons for doing so: model parameters may be transformed from a physically constrained space to an unconstrained space (the space of real numbers) to facilitate efficient sampling or numerical optimisation \citep{salvatier2016probabilistic, kucukelbir2017automatic, ardizzone2018analyzing, zhao2021bayesian, zhao2024physically}; parametrisations of data likelihood and model parameters are often modified and tested during inversion to identify optimal configurations \citep{green1995reversible, malinverno2002parsimonious, sambridge2006trans, bodin2009seismic, bodin2012transdimensionaltomo, ray2016frequency, visser2019bayesian, guo2020bayesian, biswas2022transdimensional}; different prior hypotheses are sometimes analysed and compared post Bayesian inversion \citep{michie1995machine, bishop1995neural, bernardo2002bayesian, johnson2010use, walker2014varying, papamakarios2016fast, van2021jasp, zhao2024variational, zhao2025efficient, sambridge2025trans}; the dimensionalities of model and data vectors are frequently reduced to improve the efficiency and accuracy of inversion \citep{bishop2006pattern, laloy2018training, mosser2020stochastic, aleardi2021elastic, vinciguerra2021discrete, corrales2022bayesian, scheiter2022upscaling, wang2023prior, bloem2022introducing, sun2024invertible, meles2025bayesian, li2025diffusioninv}; and subsurface properties may be parametrised based on priori physical knowledge \citep{tarantola2005inverse, wellmann20183, de2019gempy, curtis2020samples, muir2020geometric, khoshkholgh2021informed, khoshkholgh2022full, tsai2023towards, zhao2025semi, sambridge2025trans}. A fundamental assumption underlying these methodologies, whether implicit or stated explicitly, is that physical properties of reality, and our understanding of them, should not depend on the way we represent them numerically. Probability distributions that represent uncertainty about the same physical event should therefore remain invariant under changes to a different, equivalent, mathematical parametrisation. As a result, the same, \textit{physically consistent} solution to any inverse problem should be obtained if exactly the same information is considered, regardless of the parametrisations employed \citep{tarantola1982inverse, mosegaard2016inverse, mosegaard2024inconsistency}.

Unfortunately, \cite{borel1909elements} and \cite{kolmogorov1933foundations} noticed that the aforementioned assumption does not always hold. Specifically, they showed that conditional probability densities calculated over parameter subspaces can be inconsistent when evaluated under different parametrisations. This is referred to as the Borel-Kolmogorov inconsistency \cite[BK-inconsistency --][]{mosegaard2024inconsistency}: it states that even if a change of parametrisation results in consistent probability densities in the full parameter space, it can lead to inconsistent results in lower-dimensional parameter subspaces. 

While this issue might not be significant in some statistical problems in which it may be considered acceptable that a unique set of parameters is considered throughout an investigation, \cite{mosegaard2024inconsistency} demonstrated that the BK-inconsistency problem becomes critical in physical problems. Several well-established Bayesian methods, including coordinate transformation, trans-dimensional inversion, hierarchical Bayesian inversion, and Bayesian model selection methods, produce physically inconsistent solutions. Indeed, examples presented in \cite{mosegaard2024inconsistency} showed that entirely contradictory probabilistic inversion solutions can be obtained, solely because different, apparently equivalent parametrisations were considered.

This inconsistency problem might already have impacted the geoscience community. One example may lie within the global seismology community, in which different Earth models have been obtained using similar data sets, where different researchers employed different parametrisations (both in model and data space) \citep{montelli2006catalogue, houser2008shear, simmons2010gypsum, obayashi2013finite, koelemeijer2016sp12rts, lu2019tx2019slab, van2020accelerated, hosseini2020global, ritsema2020heterogeneity, cui2024glad, thrastarson2024reveal, fichtner2025high}. Unfortunately, studies seldom discuss, acknowledge, or even recognise this inconsistency in geoscientific problems, nor in other fields of physical science, so it is currently unclear to what extent it has biased results in a wide variety of fields.

In this work, we explore the consequences of the BK-inconsistency in geoscientific inference problems using synthetic and real data. We demonstrate that changing between equivalent model parametrisations can lead to significant inconsistency in conditional prior probability density functions (pdf's), despite representing the same physical prior information. The changed prior pdf's then influence posterior statistics of probabilistic inversion results, including maximum \textit{a posteriori} solutions, posterior pdf's, and subsequent uncertainty analyses, even though the solutions then incorporate the same observational data and physical prior information. Such biases certainly affect quantitative interpretation of Earth properties and structure on global, regional or local scales, as we demonstrate herein. Although for certain case-specific problems different parametrisations may yield consistent results, there is currently no generally applicable Bayesian method to avoid the BK-inconsistency. Therefore, this study indicates that careful consideration should be given when applying Bayesian inversion in situations that permit changes in parametrisation.

The rest of this paper is organised as follows. We first introduce the concept of the BK-inconsistency using a simple 2-dimensional example, followed by a mathematical explanation. We then illustrate its impact on several common geophysical inversion problems (including a real data example) before discussing the significance and implications of this work, and concluding.

\section{Theory}
\subsection{Bayesian inference}
Inverse problems often have to be solved to estimate unknown model parameter $\mathbf{m}$ given observed data $\mathbf{d}_{obs}$. In Bayesian inference, this is accomplished within a probabilistic framework, in which both unknown parameters and observations are defined as random variables. The corresponding uncertainties are characterised by probability distributions which represent an incomplete state of information \citep{tarantola2005inverse}. Bayes' theorem is then used to solve Bayesian inference (inversion) problems:
\begin{equation}
	p(\mathbf{m}|\mathbf{d}_{obs}) = \dfrac{p(\mathbf{d}_{obs}|\mathbf{m})p(\mathbf{m})}{p(\mathbf{d}_{obs})}
	\label{eq:bayes}
\end{equation}
where $p(\cdot)$ denotes a probability density function (pdf). Term $p(\mathbf{m})$ is the \textit{prior} pdf of model parameter $\mathbf{m}$ which describes prior knowledge about $\mathbf{m}$ that is independent of the data $\mathbf{d}_{obs}$. The conditional probability $p(\mathbf{d}_{obs}|\mathbf{m})$ is called the \textit{likelihood} and denotes the probability of observing data $\mathbf{d}_{obs}$ given a particular model $\mathbf{m}$ (that is, under the assumption that model $\mathbf{m}$ represents the true state of the Earth). In Bayesian inversion, this term is often calculated by introducing a \textit{forward function}, a physical relationship $\mathbf{d} = f(\mathbf{m})$ which maps the model parameter space into the data space. Term $p(\mathbf{d}_{obs})$ is called the \textit{evidence} and acts as a normalisation constant to ensure that the result of Bayesian inference is a valid probability distribution. Term $p(\mathbf{m}|\mathbf{d}_{obs})$ is the so-called \textit{posterior} probability density of model parameter $\mathbf{m}$ given data $\mathbf{d}_{obs}$ and taking account of prior information in $p(\mathbf{m})$; the posterior distribution describes the complete, probabilistic solution to the inverse problem.

Both Monte Carlo and variational inference methods are widely employed to solve Bayesian geoscientific inference problems. For all numerical examples presented in this study, we use Monte Carlo sampling methods to explore and characterise the posterior distribution, using a very large number of samples to ensure adequate convergence of the algorithm and to remove the possibility that inaccurate evaluation of probability distributions or variational approximations could bias our analyses.

\subsection{Borel-Kolmogorov inconsistency}
\label{section:toy}
Following \cite{mosegaard2024inconsistency}, the key research question we aim to investigate is: can we define a conditional probability distribution of the same physical event \textit{uniquely} under different parametrisations? Specifically, say we have a prior pdf $p(\mathbf{m})$ of $\mathbf{m}$. Given a lower dimensional parameter subspace defined by a condition $c$, will the conditional pdf $p(\mathbf{m}|c)$ remain physically consistent if evaluated using different parametrisations of $\mathbf{m}$? 

\begin{figure}
	\centering\includegraphics[width=\textwidth]{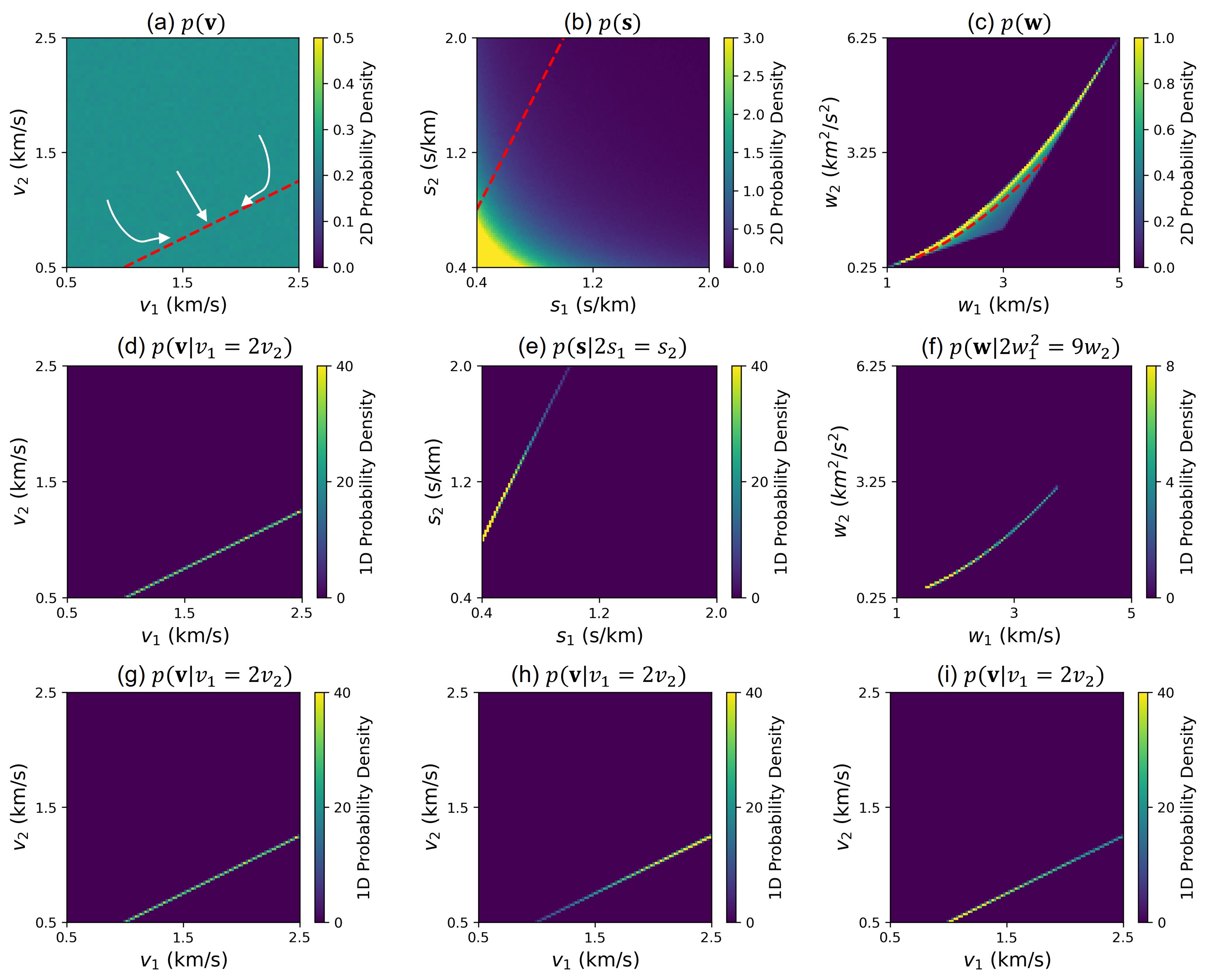}
	\caption{Probability density functions (pdf's) of the same 2D uniform prior distribution defined for a velocity random variable $\mathbf{v}$, represented by three different parametrisations: (a) velocity $\mathbf{v}$, (b) slowness $\mathbf{s}$, and (c) $\mathbf{w}$. Dashed red lines illustrate a parameter subspace manifold defined by a condition $v_1 = 2v_2$ (and equivalently $2s_1 = s_2$; $2w_1^2 = 9w_2$). White arrows in (a) exemplify different ways to approach the 1D conditional subspace (dashed red line) in the 2D full parameter space. (d), (e) and (f) Conditional pdf's displayed in the full (2D) space. (g), (h) and (i) Condition pdf's using the velocity parametrisation $\mathbf{v}$, obtained by transforming conditional pdf's with different parametrisations in (d), (e) and (f) to $p(\mathbf{v})$, using the change of variable formulae in equations \ref{eq:s_from_v} and \ref{eq:w_from_v}. Panels (d) and (g) are identical. Note that these conditional pdf's have zero probability in the full space, so are displayed on the 2D plots for illustration only.}
	\label{fig:2d_v1eq2v2}
\end{figure}

Similarly to \cite{mosegaard2024inconsistency}, we start with a simple 2-dimensional (2D) example. Consider a random variable $\mathbf{v}$ which represents a 2D seismic velocity model vector $\mathbf{v}^T = [v_1, v_2]^T$. Define a uniform prior distribution for $\mathbf{v} \in [0.5\ km/s, 2.5\ km/s]^2$. The prior probability value is constant with $p(\mathbf{v}) = k = 0.25$ inside of the prior bounds (otherwise $p(\mathbf{v})=0$), as displayed in Figure \ref{fig:2d_v1eq2v2}a. In the rest of this paper we only focus on the non-zero sections of each probability distribution, implicitly ignoring sets of parameter values with zero probability.

We define two additional parametrisations to represent the same prior information: slowness model vector $\mathbf{s}^T = [s_1, s_2]^T = [1/v_1, 1/v_2]^T$ and another vector $\mathbf{w}^T = [w_1, w_2]^T = [v_1 + v_2, v_1v_2]^T$. Both slowness and velocity parametrisations, which are reciprocal to each other (i.e., $v = 1/s$), are used commonly in seismology. Similar reciprocal parameter pairs applied in other fields of geophysics include period $T$ and frequency $f=1/T$, resistivity $r$ and conductivity $\sigma = 1/r$, and the stiffness and compliance tensors in Hooke's law \citep{mosegaard2016inverse}. While the $\mathbf{w}$ parametrisation may at first appear less familiar, it combines a product and a sum of variables which are common re-parametrisations: in seismic imaging, physical parameters density $\rho$ and velocity $v$ are often combined as acoustic impedance $I = \rho v$, and the arithmetic average of any two physical properties is simply their sum divided by two. 

According to the change of variable formula, probability densities $p(\mathbf{s})$ and $p(\mathbf{w})$ are related to $p(\mathbf{v})$
\begin{equation}
	p(\mathbf{s}) = p(\mathbf{v})\left|\det \dfrac{\partial \mathbf{s}}{\partial \mathbf{v}}\right|^{-1} = k s_1^{-2} s_2^{-2}~~, \quad \text{where} \quad \dfrac{\partial \mathbf{s}}{\partial \mathbf{v}} = \left[{\begin{array}{cc}
			-v_1^{-2} & 0 \\
			0 & -v_2^{-2} \\
	\end{array}}\right]
	\label{eq:s_from_v}
\end{equation}
and
\begin{equation}
	p(\mathbf{w}) = p(\mathbf{v})\left|\det \dfrac{\partial \mathbf{w}}{\partial \mathbf{v}}\right|^{-1} = \dfrac{k}{\sqrt{w_1^2-4w_2}}~~, \quad \text{where} \quad \dfrac{\partial \mathbf{w}}{\partial \mathbf{v}} = \left[{\begin{array}{cc}
			1 & 1 \\
			v_2 & v_1 \\
	\end{array}}\right]
	\label{eq:w_from_v}
\end{equation}
where symbol $|\cdot|$ denotes the absolute value and $\det(\cdot)$ is the determinant of a matrix \citep{rezende2015variational}. In this particular example constant $k=0.25$. Figures \ref{fig:2d_v1eq2v2}b and \ref{fig:2d_v1eq2v2}c display the non-zero sections of $p(\mathbf{s})$ and $p(\mathbf{w})$, respectively, and the three pdf's in Figures \ref{fig:2d_v1eq2v2}a, \ref{fig:2d_v1eq2v2}b and \ref{fig:2d_v1eq2v2}c represent the same prior information in full (2D) parameter space. In other words, if we were to reverse-transform $p(\mathbf{s})$ and $p(\mathbf{w})$ back to $p(\mathbf{v})$ by rearranging equations \ref{eq:s_from_v} and \ref{eq:w_from_v}, we would obtain two uniform pdf's that are identical to that shown in Figure \ref{fig:2d_v1eq2v2}a. In practice, these different parametrisations might be employed by different people based on their field of expertise and preferences, yet they describe the same physical information.

We now introduce an additional constraint, $v_1 = 2v_2$ (and equivalently, $2s_1 = s_2$ and $2w_1^2 = 9w_2$), which creates a 1-dimensional (1D) parameter subspace \citep{mosegaard2024inconsistency}. The manifold of this conditional subspace is illustrated by dashed red lines in Figures \ref{fig:2d_v1eq2v2}a, \ref{fig:2d_v1eq2v2}b and \ref{fig:2d_v1eq2v2}c, respectively, which again represent exactly the same information. Figures \ref{fig:2d_v1eq2v2}d, \ref{fig:2d_v1eq2v2}e and \ref{fig:2d_v1eq2v2}f show the corresponding 1D conditional probability densities computed using different parametrisations, displayed in the 2D space. The values of these conditional distributions are just the values of the 2D distributions on the red lines in the corresponding panels above, normalised to have integral one along the line to ensure that they represent valid probability distributions. Note that these densities are only displayed in 2D for visualisation purposes, these 1D conditional pdf's have zero probability if they are considered within the 2D space since the 1D red line has zero width in 2D.

We can transform the conditional pdf's in Figures \ref{fig:2d_v1eq2v2}e and \ref{fig:2d_v1eq2v2}f back to the velocity parametrisation using equations \ref{eq:s_from_v} and \ref{eq:w_from_v}. Surprisingly, we obtain the results displayed in panels \ref{fig:2d_v1eq2v2}h and \ref{fig:2d_v1eq2v2}i, respectively. Despite defining a consistent 2D uniform probability distribution across all three parametrisations, and applying exactly the same condition to each 2D parameter space, we obtain inconsistent conditional distributions along the 1D parameter subspace (manifold). A uniform conditional pdf appears in Figure \ref{fig:2d_v1eq2v2}g using the (original) velocity parametrisation, whereas the probabilities in Figures \ref{fig:2d_v1eq2v2}h and \ref{fig:2d_v1eq2v2}i (as well as the middle and right panels in Figure \ref{fig:2d_3conditions}a) are biased towards higher and lower velocity values, respectively. Figure \ref{fig:2d_3conditions_1d_marginals}a displays the marginal pdf's of $v_1$ for these three cases, further highlighting the inconsistency of these results.

This phenomenon of inconsistency, known as the \textit{Borel-Kolmogorov paradox}, was discussed by \cite{borel1909elements} and \cite{kolmogorov1933foundations}. \cite{mosegaard2024inconsistency} investigated its consequences systematically in general Bayesian inference problems, and referred to it as the \textit{BK-inconsistency}. It states that changes to equivalent parametrisations ensure consistent probability densities in full parameter space, but lead to inconsistent results in lower-dimensional parameter subspaces. Intuitively, this can be understood as follows. When we define a lower-dimensional subspace from the original high-dimensional space, we implicitly need to consider the limit of taking smaller and smaller volumes around the subspace in question (the red line in Figure \ref{fig:2d_v1eq2v2}a). As we approach the limit of zero volume in different ways, we will observe different limits: for example, approaching perpendicularly to the line (the central white arrow), we would consider increasingly tiny squares around each point on the line, whereas approaching diagonally (other white arrows) we would consider parallelograms. The squares and parallelograms will contain different probability masses if the 2D distribution is non-uniform, and so their limiting values at zero volume will differ.

\subsection{Mathematical explanation}
\label{section:math}

Under the velocity parametrisation, the non-zero section of the conditional distribution is homogeneous within the parameter subspace. As a function of $v_1$,
\begin{equation}
	p(v_1\ |\ [v_1=2v_2]) = k_1
	\label{eq:manifold_v1eq2v2}
\end{equation}
where $k_1$ is a constant. Similarly, we can apply this condition to $p(\mathbf{s})$ and $p(\mathbf{w})$ by substituting $2s_1 = s_2$ and $2w_1^2 = 9w_2$ into equations \ref{eq:s_from_v} and \ref{eq:w_from_v}, as a function of $s_1$ and $w_1$, respectively:
\begin{equation}
	p(s_1\ |\ [2s_1=s_2]) = k_2s_1^{-4}\quad , \quad p(w_1\ |\ [2w_1^2 = 9w_2]) = k_3w_1^{-1}
	\label{eq:manifold_s_w}
\end{equation}
where $k_2$ and $k_3$ are normalisation constants. Transforming these two conditional pdf's back to the velocity space gives (according to the change of variable rule)
\begin{equation}
	\begin{split}
		p(v_1\ |\ [v_1=2v_2]) &= p(s_1\ |\ [2s_1=s_2]) \left|\dfrac{d s_1}{d v_1} \right| = k_2' v_1^2 \\
		p(v_1\ |\ [v_1=2v_2]) &= p(w_1\ |\ [2w_1^2 = 9w_2]) \left|\dfrac{d w_1}{d v_1} \right| = k_3'v_1^{-1}
	\end{split}
	\label{eq:manifold_s_w_tov}
\end{equation}
which are inconsistent with equation \ref{eq:manifold_v1eq2v2}. Equation \ref{eq:manifold_s_w_tov} states that the conditional pdf is biased towards high velocity values under the slowness parametrisation and towards low velocity values under the $\mathbf{w}$ parametrisation, which align with the results in Figures \ref{fig:2d_v1eq2v2}h and \ref{fig:2d_v1eq2v2}i, respectively.

The problem concealed within the above derivation is that the conditional pdf's in equations \ref{eq:manifold_v1eq2v2} and \ref{eq:manifold_s_w} have zero probability in the original full (2D) space. In other words, the conditional pdf's in Figures \ref{fig:2d_v1eq2v2}d, \ref{fig:2d_v1eq2v2}e and \ref{fig:2d_v1eq2v2}f have zero hyper-volume (area in this 2D scenario) in the full parameter space, and hence contain zero probability at any point. The change of variable rule ensures correct transformation of pdf's across different parametrisations in full parameter space, as illustrated in Figures \ref{fig:2d_v1eq2v2}a, \ref{fig:2d_v1eq2v2}b and \ref{fig:2d_v1eq2v2}c where exactly the same prior information in the 2D space is described. Inconsistencies arise when we constrain our attention to parameter \textit{subspaces} under different parametrisations. More specifically, this example illustrates that one can not assign values from a higher-dimensional density to a lower-dimensional one at co-located points \citep{mosegaard2024inconsistency}. This is easy to understand from a Bayesian perspective: all information is represented by random variables and their probability distributions. In this example, we only define physical information for 2D velocity values. No information is defined directly within the 1D subspace.

To summarise, this example verifies that exactly the same physical events defined in a high-dimensional space can and do incur inconsistent probabilistic results in lower-dimensional subspaces (manifolds) when evaluated under different parametrisations. Note that the BK-inconsistency is specific to continuous probability distributions (parameter spaces). It does not exist for discrete probability mass functions (discrete random variables), since the limiting process required to approach a parameter subspace continuously is not necessary (or even possible) in discrete settings.

\subsection{BK-inconsistency with different conditions}
\begin{figure}
	\centering\includegraphics[width=\textwidth]{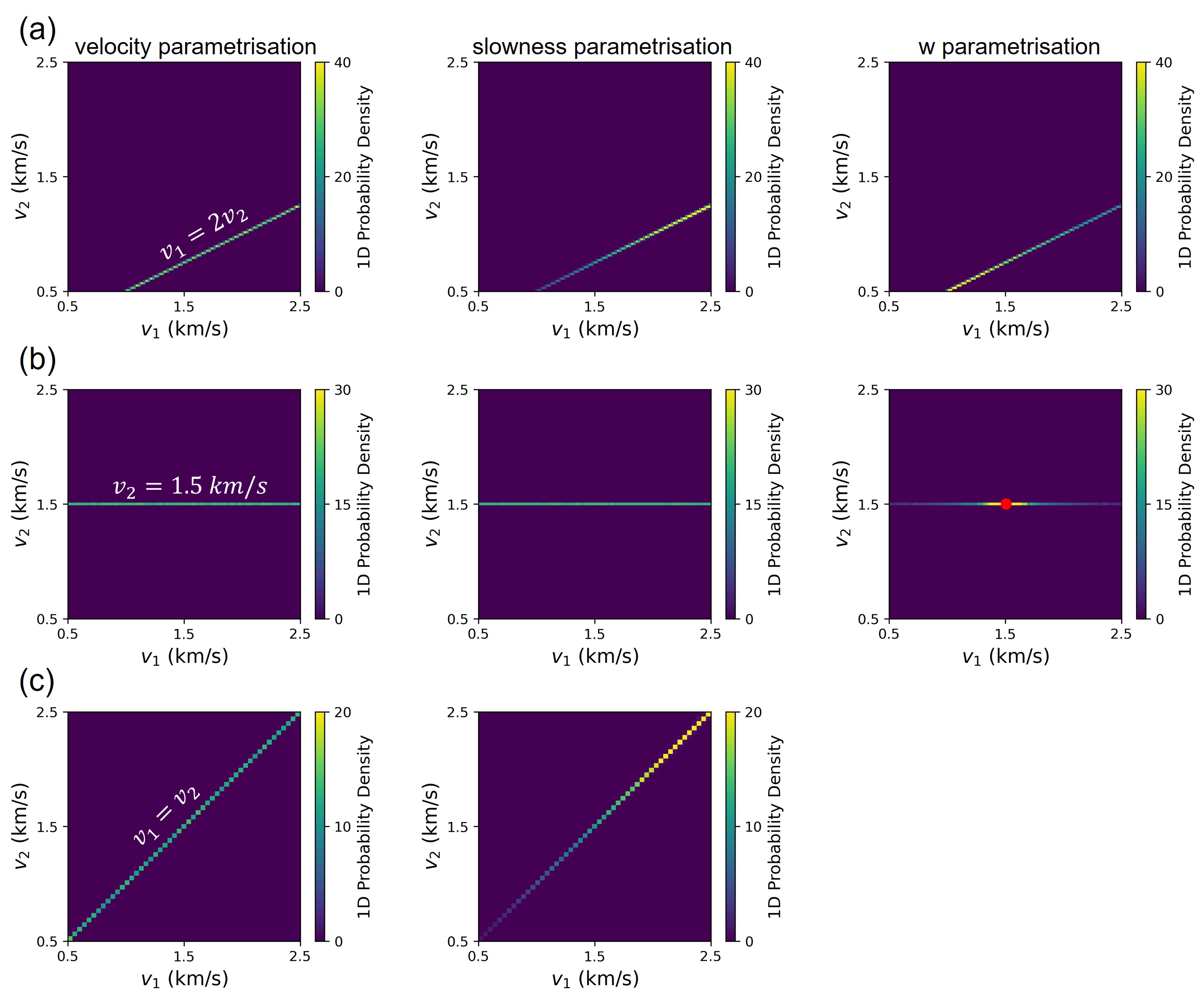}
	\caption{Conditional pdf's of velocity parameters along manifold subspace: (a) $v_1=2v_2$, (b) $v_2=1.5\ km/s$, and (c) $v_1=v_2$. From left to right, each column displays the results obtained using one particular parametrisation. Note that $p(\mathbf{w})$ does not have valid definition given $v_1=v_2$, therefore we leave this panel blank. Red dot in the right panel in (b) denotes a singular point of $p(\mathbf{w})$ given $v_1=v_2=1.5\ km/s$.}
	\label{fig:2d_3conditions}
\end{figure}

\begin{figure}
	\centering\includegraphics[width=\textwidth]{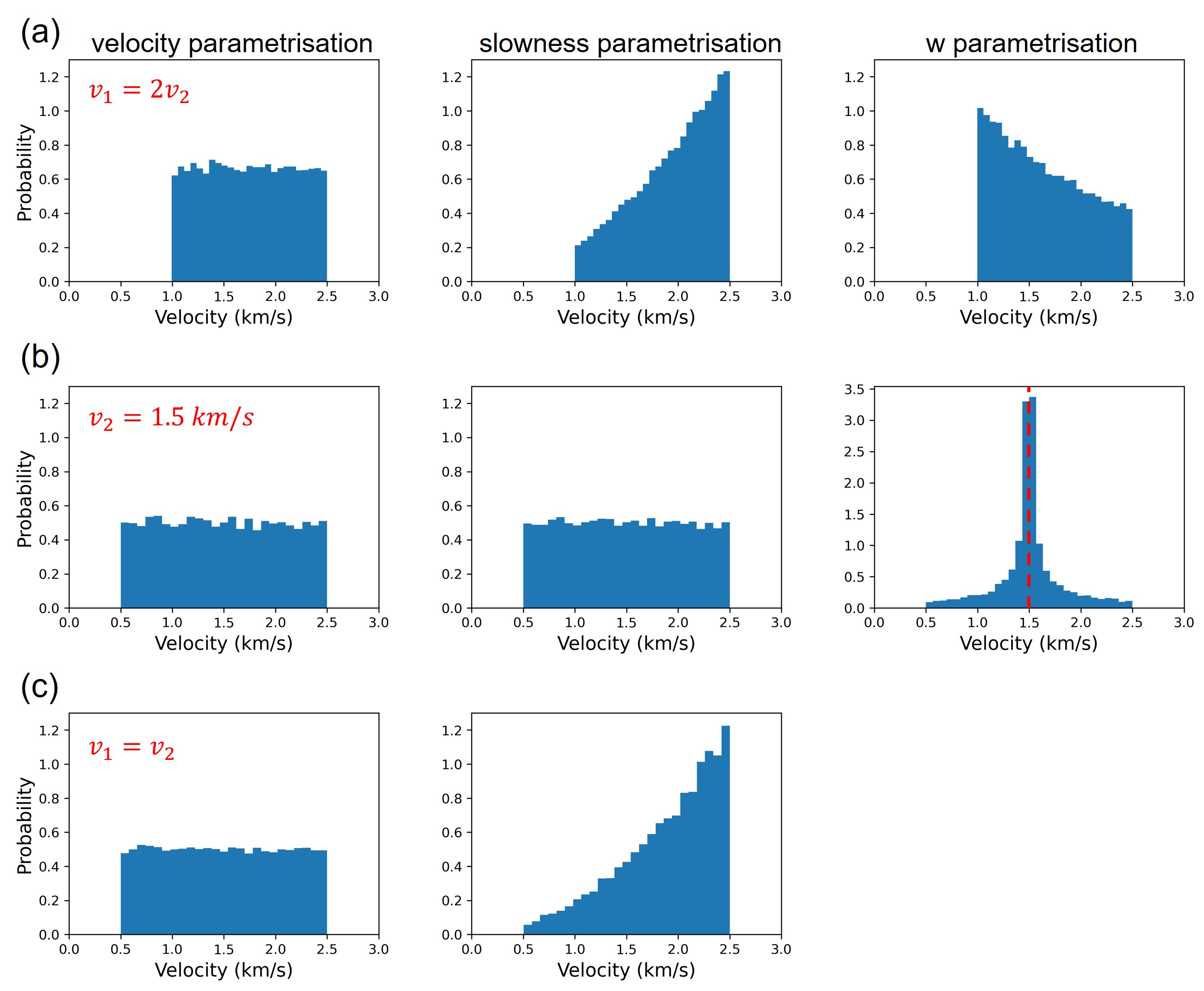}
	\caption{Conditional marginal pdf's of parameter $v_1$, corresponding to the results shown in Figure \ref{fig:2d_3conditions}. Dashed red line ($v_1=1.5\ km/s$) in the right panel in (b) represents a singular point of $p(\mathbf{w})$.}
	\label{fig:2d_3conditions_1d_marginals}
\end{figure}

Consider the BK-inconsistency under two alternative parameter conditions, or subspaces: $v_2 = 1.5\ km/s$ ($s_2 = 2/3\ s/km$; $w_1 = 1.5 + 2w_2/3$) and $v_1 = v_2$ ($s_1 = s_2$; $w_1^2 = 4w_2$). From top to bottom row, Figures \ref{fig:2d_3conditions}a, \ref{fig:2d_3conditions}b and \ref{fig:2d_3conditions}c display the obtained pdf's of the velocity value $\mathbf{v}$ conditioned on different constraints denoted in the left panel in each figure. Again, we display the 1D conditional density distributions on 2D plots only to place the marginal 1D distributions in context, even though they would have zero probability value in the full (2D) space. From left to right, each column shows the results obtained using the three parametrisations, respectively. The three panels in Figure \ref{fig:2d_3conditions}a correspond to those in Figures \ref{fig:2d_v1eq2v2}g, \ref{fig:2d_v1eq2v2}h and \ref{fig:2d_v1eq2v2}i. Figure \ref{fig:2d_3conditions_1d_marginals} shows the corresponding 1D marginal pdf's of parameter $v_1$. 

Note that the results conditioned on $v_1=v_2$ for the $\mathbf{w}$ parametrisation do not exist (hence the panel in the third row and third column in both Figures \ref{fig:2d_3conditions} and \ref{fig:2d_3conditions_1d_marginals} is blank). This is because the determinant of the Jacobian matrix $\partial \mathbf{w}/\partial \mathbf{v}=0$ given $v_1=v_2$, meaning that there is no valid definition for $p(\mathbf{w})$ along the manifold $v_1=v_2$ based on $p(\mathbf{v})$ and the change of variable rule (equation \ref{eq:w_from_v}).

We can draw the following conclusions from Figures \ref{fig:2d_3conditions} and \ref{fig:2d_3conditions_1d_marginals}. First, from each row we find that different parametrisations result in different conditional pdf's, despite the fact that the same information is injected. Second, the results across different columns show that the magnitude of inconsistency varies with conditions (subspace manifolds) applied. For example, under the slowness parametrisation, the conditional pdf's are biased towards high velocities for $v_1=v_2$ and $v_1=2v_2$; however, the results appear consistent given $v_2 = 1.5\ km/s$. For the $\mathbf{w}$ parametrisation, the results are biased towards low velocities for $v_1=2v_2$. For $v_2 = 1.5\ km/s$, the conditional pdf is biased towards $v_1=1.5\ km/s$. This is because $p(\mathbf{w})$ reaches a singularity at $v_1=1.5\ km/s$ (denoted by a red dot in Figure \ref{fig:2d_3conditions}b and a dashed red line in Figure \ref{fig:2d_3conditions_1d_marginals}b), given the condition $v_2=1.5\ km/s$. According to equation \ref{eq:w_from_v}, when $v_1 \rightarrow 1.5$, $p(\mathbf{w}) \rightarrow \infty$. Third, the left two panels in Figures \ref{fig:2d_3conditions}b and \ref{fig:2d_3conditions_1d_marginals}b demonstrate that while some cases may show consistent results when comparing conditional pdf's evaluated from different parametrisations, this does not negate the presence of the BK-inconsistency (the right hand panels differ). Instead, it shows that identical inconsistencies can sometimes manifest across different parametrisations.

\section{BK-inconsistency in geoscientific inference problems}
In the following we provide several examples that demonstrate the consequences of ignoring the BK-inconsistency in geoscientific inference (inverse) problems: seismic impedance inversion, surface wave dispersion inversion and travel time tomography.

\subsection{Seismic impedance inversion}
We first consider a simple post-stack seismic impedance inversion problem using a laterally homogeneous layered model mirroring practical studies \citep{lindseth1979synthetic, russell1988introduction, alves2014simulation, maurya2016comparison, maurya2018comparing}. Within each layer, we represent subsurface properties using 2 parameters: seismic P-wave velocity $v$ and density $\rho$, and define a uniform prior pdf for these two parameters: $[v,\rho] \in [1,3]\ (km/s) \times [1,3]\ (g/cm^3)$. We consider another parametrisation to represent exactly the same uniform prior information, in which subsurface properties are described by P-wave velocity $v$ and impedance $I$ with the relation $I=v\rho$. Similarly to equation \ref{eq:w_from_v}, we have
\begin{equation}
	p([v,I]^T) = p([v,\rho]^T)\left|\det \dfrac{\partial [v,I]^T}{\partial [v,\rho]^T}\right|^{-1} = \dfrac{k'}{v}~~, \quad \text{where} \quad \dfrac{\partial [v,I]^T}{\partial [v,\rho]^T} = \left[{\begin{array}{cc}
		1 & 0 \\
		\rho & v \\
	\end{array}}\right]
	\label{eq:I_from_v}
\end{equation}
where $k'$ is a constant. Both parametrisations are commonly used in seismic impedance inversion, the former by \cite{maurya2018comparing} and the latter by \cite{alves2014simulation}, for example. 

Now we introduce an additional condition (constraint) between $v$ and $\rho$:
\begin{equation}
	\rho = 0.31(1000v)^{0.25}
	\label{eq:gardner}
\end{equation}
where $v$ and $\rho$ are given in $km/s$ and $g/cm^3$, respectively, aligning with the units of their prior bounds. This is referred to as Gardner's empirical equation which is commonly used in the exploration and near-surface reflection seismology community \citep{gardner1974formation}. 

\begin{figure}
	\centering\includegraphics[width=\textwidth]{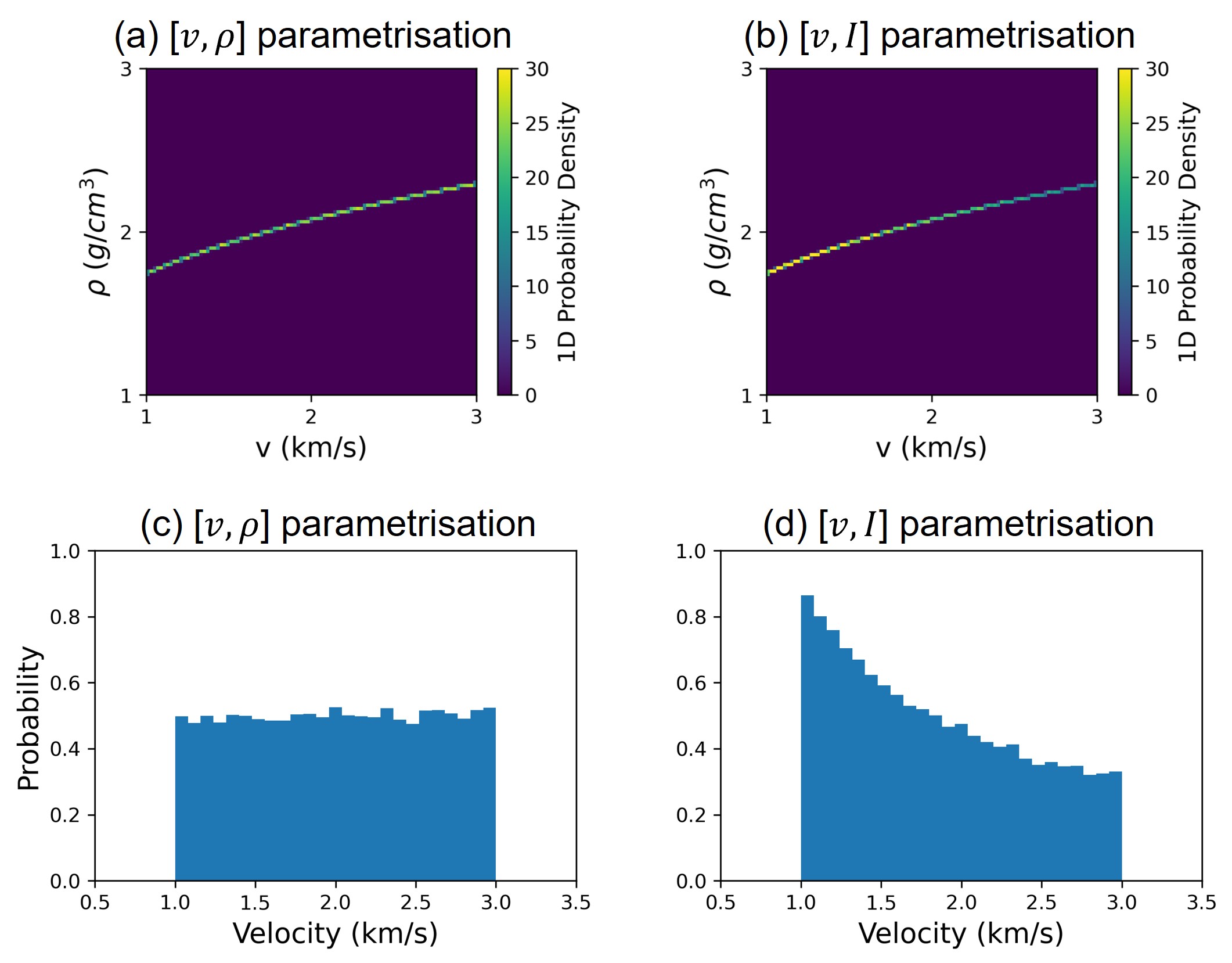}
	\caption{Prior pdf's evaluated under (a) $[v,\rho]$ parametrisation and (b) $[v,I]$ parametrisation, conditioned on the Gardner's empirical relation in equation \ref{eq:gardner}. (c) and (d) are the corresponding marginal pdf's of P-wave velocity $v$.}
	\label{fig:impedance_prior}
\end{figure}

Given the above setup, we have a conditional inter-parameter relationship which defines a parameter subspace. The BK-inconsistency therefore exists between conditional pdf's evaluated under the two parametrisations. Figures \ref{fig:impedance_prior}a and \ref{fig:impedance_prior}b show the pdf's of $[v,\rho]$ conditioned on the relationship in equation \ref{eq:gardner}, and evaluated under the two parametrisations (1D conditional pdf's are displayed in 2D space for illustration). Figures \ref{fig:impedance_prior}c and \ref{fig:impedance_prior}d display the corresponding marginal pdf's of P-wave velocity $v$. As expected, we observe clear evidence of inconsistent prior pdf's between these two parametrisations, despite the injection of identical prior information into the full (2D) parameter space in each case.

\subsubsection{Inverting parameters in a single layer}
Now we investigate the impact of the inconsistent prior information depicted in Figure \ref{fig:impedance_prior} on Bayesian posterior solutions. We use a laterally homogeneous 2-layered model as the true model, with parameters defined for these layers as follows: in layer 1, $v_1 = 1.5\ km/s$ and $\rho_1 = 1.2\ g/cm^3$, and in layer 2, $v_2 = 2.7\ km/s$ and $\rho_2 = 2.2\ g/cm^3$. All of these values fall within the non-zero section of the uniform prior distribution defined above. We consider a post-stack seismic impedance inversion scenario, in which we assume that we can observe or estimate the seismic reflection coefficient of the interface between the layers. The forward function that estimates the seismic reflection coefficient $r$ for waves arriving at normal incidence to the surface, given the subsurface properties, can be expressed as
\begin{equation}
	r = \dfrac{I_2 - I_1}{I_2 + I_1} = \dfrac{v_2\rho_2 - v_1\rho_1}{v_2\rho_2 + v_1\rho_1}
	\label{eq:reflectivity}
\end{equation}

We start with an overly simplified example in which we assume that the true values of parameters in the first layer are known so we only invert for parameters $[v_2,\rho_2]$ in the second layer in a 2-dimensional inversion. The observed reflectivity datum is 0.535, and a Gaussian data uncertainty with an observational uncertainty (standard deviation) of 0.14 is used to define the likelihood function. We employ the uniform prior distribution defined above, and apply Gardner's relation between $v_2$ and $\rho_2$. The inversion is solved by the Metropolis-Hastings Markov chain Monte Carlo algorithm \cite[MH-McMC --][]{mosegaard1995monte} under the two parametrisations. For the $[v,I]$ parametrisation, the prior probability value of any proposed sample is calculated using equation \ref{eq:I_from_v}. The corresponding inversion (sampling) results are a set of posterior samples of $[v_2,I_2]$ values, which are transformed back to $[v_2,\rho_2]$ for comparison.

\begin{figure}
	\centering\includegraphics[width=\textwidth]{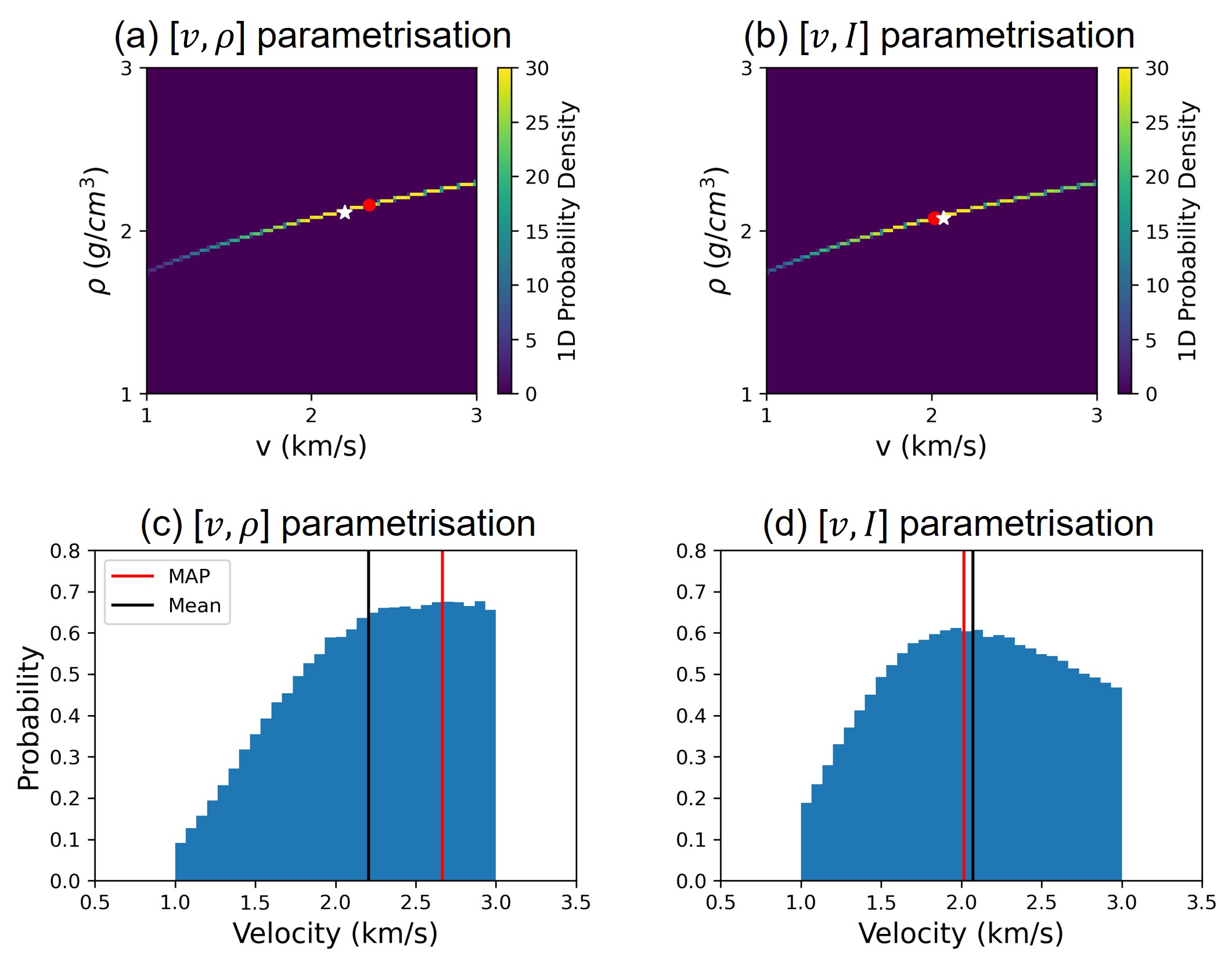}
	\caption{Conditional posterior pdf's of $[v,\rho]$ parameters in the bottom layer, given fixed values in the top layer, and a single reflectivity datum. Key as in Figure \ref{fig:impedance_prior}. Red dots and white stars in panels (a) and (b) mark respectively the maximum \textit{a posteriori} (MAP) and mean solutions obtained under each parametrisation.}
	\label{fig:impedance_posterior_1layer}
\end{figure}

Figure \ref{fig:impedance_posterior_1layer} displays the posterior pdf's (after the latter transformation to $[v_2,\rho_2]$) obtained using the two parametrisations. Each panel corresponds to the respective panel in Figure \ref{fig:impedance_prior}. Red dots in Figures \ref{fig:impedance_posterior_1layer}a and \ref{fig:impedance_posterior_1layer}b, along with red lines in Figures \ref{fig:impedance_posterior_1layer}c and \ref{fig:impedance_posterior_1layer}d, highlight the maximum \textit{a posteriori} (MAP) or the most likely solutions found using the different parametrisations. White stars and black lines highlight the corresponding mean solutions. 

Using identical \textit{a priori} information and observational datum, we obtain different posterior pdf's, and significantly different MAP and mean solutions. These differences arise solely from the changes in prior pdf's evaluated using different parametrisations in Figure \ref{fig:impedance_prior} -- a consequence of the BK-inconsistency. Specifically, since the prior pdf of the velocity value evaluated using the $[v,I]$ parametrisation is biased towards the low velocity direction, so are the corresponding posterior pdf and the MAP and mean solutions. 

\subsubsection{Inverting parameters in both layers simultaneously}
\begin{figure}
	\centering\includegraphics[width=\textwidth]{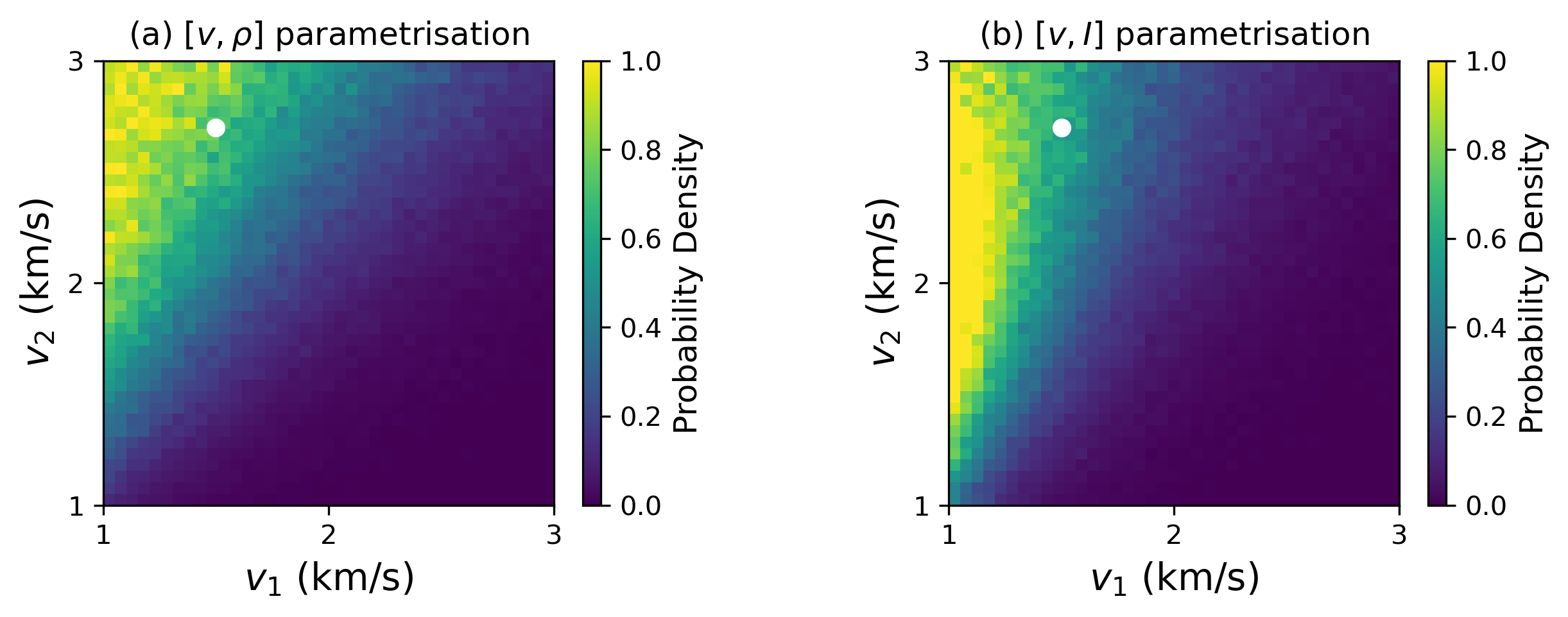}
	\caption{Posterior bivariate pdf's of two velocity parameters $v_1$ and $v_2$, each obtained using the different parametrisations noted in the title. White dots denote true velocity values in these two layers.}
	\label{fig:impedance_posterior_2layers_v1v2}
\end{figure}

\begin{figure}
\centering\includegraphics[width=\textwidth]{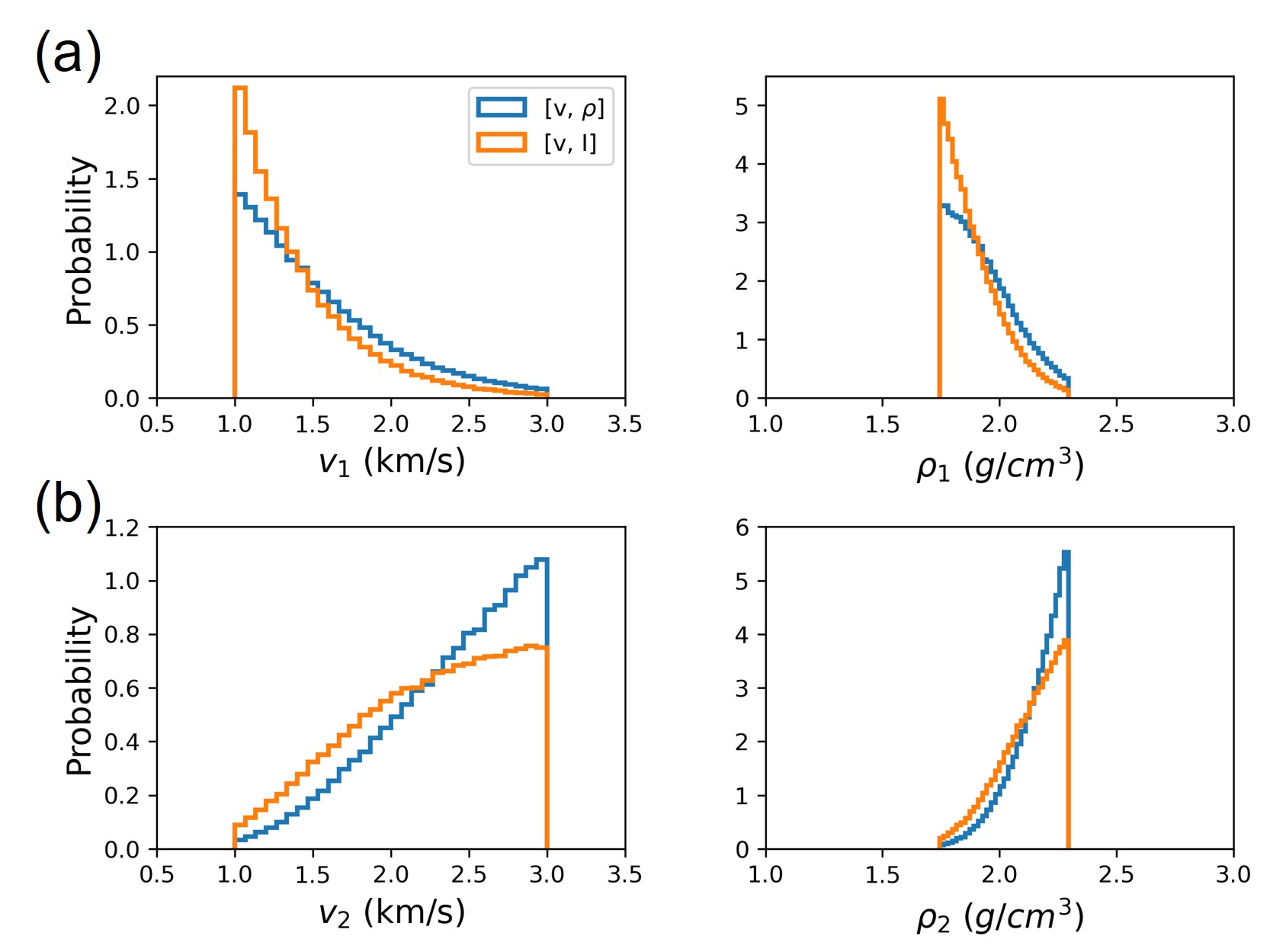}
\caption{Posterior marginal pdf's obtained using two different parametrisations (see legend) for the 4-dimensional reflectivity inversion problem. (a) and (b) represent marginal pdf's of parameters in the first and second layers, respectively.}
\label{fig:impedance_posterior_2layers_marginals}
\end{figure}

We now invert for the values of parameters in both layers simultaneously in a 4-dimensional inversion (for $v_1, v_2, \rho_1$ and $\rho_2$). Other experimental details are the same as those in the previous test. Gardner's relation (equation \ref{eq:gardner}) is applied to constrain velocity and density values within each layer, and MH-McMC is employed for inversion using each parametrisation. Figure \ref{fig:impedance_posterior_2layers_v1v2} displays the posterior bivariate marginal pdf's of $[v_1, v_2]$ in the two layers, evaluated using each parametrisation. A white dot in each panel indicates the true velocity value. Figures \ref{fig:impedance_posterior_2layers_marginals}a and \ref{fig:impedance_posterior_2layers_marginals}b show the posterior marginal histograms of velocity and density values in the two layers. Similarly to the previous test, different posterior solutions are obtained due to the inconsistent conditional prior pdf's.

These two tests demonstrate the consequences of ignoring the BK-inconsistency in solving inverse problems: if we consider only one of the two parametrisations we would end up with potentially biased probabilistic results, leading to incorrect interpretation of subsurface properties and poor consequent decision-making. Clearly we could also construct a third parametrisation that might lead to different results again. How are we to discriminate which set of results should be interpreted, or how these different sets of results should be combined? This example suggests a partial explanation of why different groups using different parametrisations are likely to arrive at different solutions for the MAP, mean or posterior distribution of inverse problems. While the difference in posterior pdf's is small in this particular example, below we show that they are far larger in other problems. 

\subsection{Surface wave dispersion inversion}
\label{sec:swi}

In this example, we investigate the BK-inconsistency in seismic surface wave dispersion inversion using a real dataset. Dispersion inversion is a technique used extensively to image the Earth's interior structures in both global and regional seismology. In surface wave inversion, we estimate subsurface shear wave velocity profiles over depth using surface wave dispersion curves (their velocity variations with frequency of oscillation), since surface wave dispersion is primarily sensitive to shear wave velocity variations \citep{aki2002quantitative}. However, in dispersion inversion, the forward function involves modelling a dispersion curve given a layered Earth model represented by depth profiles of all of P-wave velocity $v_p$, shear wave velocity $v_s$ and density $\rho$. Within each layer, the three parameters are often linked by empirical relations \citep{trampert1995global, curtis1998eurasian, devilee1999efficient, shapiro2002monte, aki2002quantitative, simons2002multimode, xia2003inversion, wathelet2004surface, yao2006surface, meier2007fully, de2011ambient, bodin2012transdispersion, haney2015nonperturbational, galetti2017transdimensional, zhang20201, fone2024ambient, magrini2025bayesbay}. For example, $v_p$ is typically derived from $v_s$ using a fixed $v_p/v_s$ ratio, often taken to be $v_p = 1.73v_s$, and $\rho$ can be calculated from $v_p$ using Gardner's equation (equation \ref{eq:gardner}). Despite the relevance of all three parameters, this allows seismologists to focus on $v_s$ profiles, so parameters $v_p$ and $\rho$ become nuisance parameters in the forward function.

We define three parametrisations to represent the above 3-parameter system within each layer, all of which have been used in the literature: (1) $[v_p, v_s, \rho]$, (2) $[s_p, s_s, \rho] = [1/v_p, 1/v_s, \rho]$, and (3) $[v_p, v_s, I] = [v_p, v_s, v_p\rho]$. By analysing the effects of these parametrisations within a conditional subspace defined by the above inter-parameter relations, we demonstrate the inconsistency of results in the evaluated posterior probability solutions. According to Figures \ref{fig:2d_v1eq2v2} and \ref{fig:impedance_prior}, $[s_p, s_s, \rho]$ and $[v_p, v_s, I]$ parametrisations bias the prior probability density of $v_s$ towards high and low velocity directions, respectively, even though the latter only re-parametrises nuisance parameters of the forward problem.

\begin{figure}
	\centering\includegraphics[width=\textwidth]{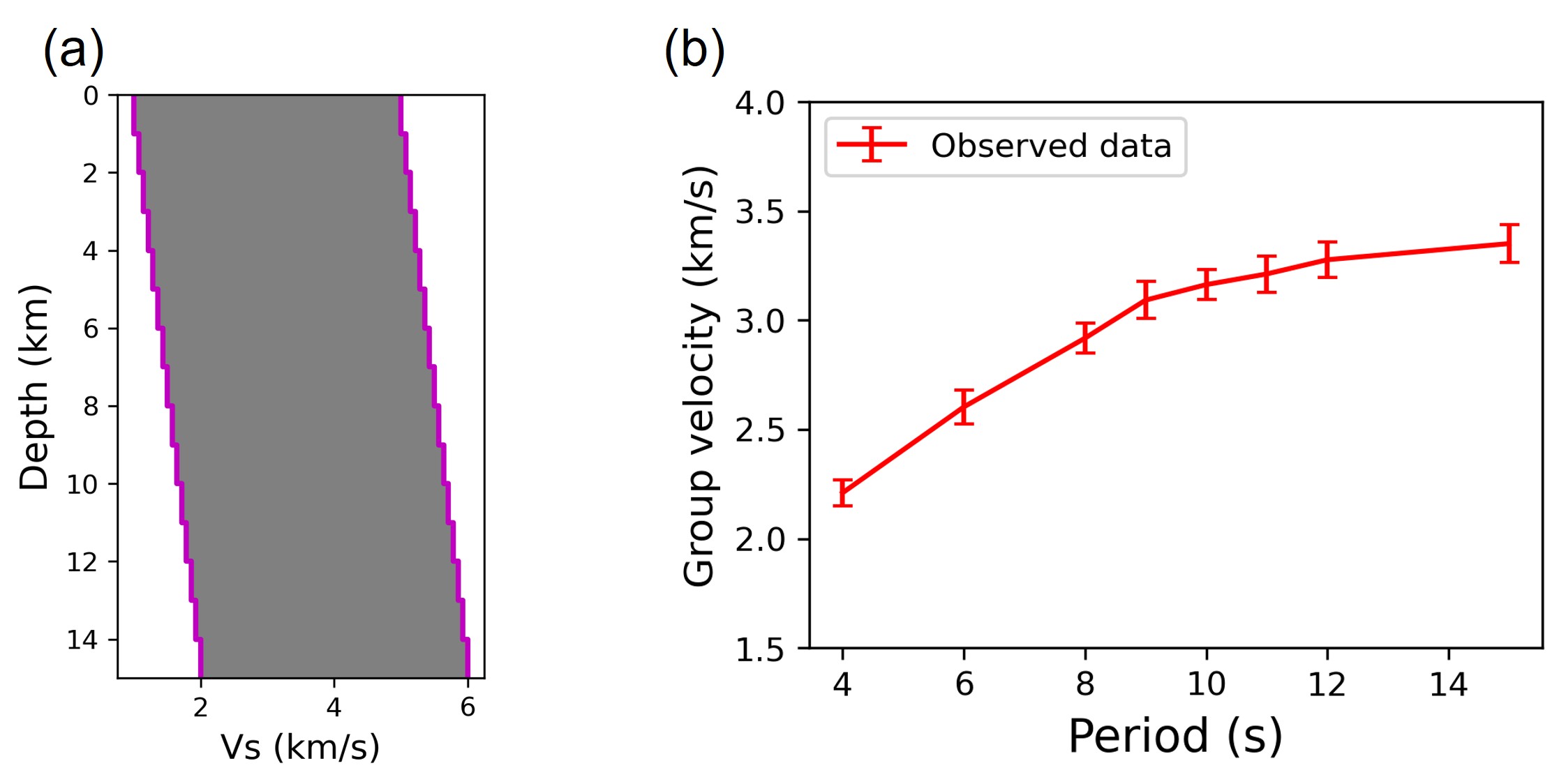}
	\caption{(a) Parametrisation of 15 subsurface layers in the surface wave inversion example. Magenta lines show lower and upper bounds of a uniform prior distribution of $v_s$ in each layer. (b) Observed Love wave group velocity dispersion curve at geographical location 55.93$^\circ$N latitude and 3.18$^\circ$W longitude, near Edinburgh city centre. Error bars show data standard deviations used to define an uncorrelated Gaussian likelihood function.}
	\label{fig:swi_truevs_data}
\end{figure}

We perform a dispersion inversion using real data to demonstrate the effect of the BK-inconsistency. The dispersion curve inverted is from geographical location 55.93$^\circ$N latitude and 3.18$^\circ$W longitude near Edinburgh city centre, obtained from ambient noise Love wave tomography of the British Isles at periods of 4 s, 6 s, 8 s, 9 s, 10 s, 11 s, 12 s, and 15 s \citep{galetti2017transdimensional, zhao2022interrogating}, as shown by a red curve in Figure \ref{fig:swi_truevs_data}b. We use this data to invert for the 1D shear wave velocity structure beneath this location down to 15 km depth. In Figure \ref{fig:swi_truevs_data}a, we discretise the subsurface structure into 15 laterally homogeneous layers. The top 14 each have a thickness of 1 km and the last layer extends to infinite depth to represent a half space. We define uniform prior distributions for $v_p$, $v_s$ and $\rho$ parameters, respectively. Lower and upper bounds for the $v_s$ model are denoted by magenta lines in Figure \ref{fig:swi_truevs_data}a. These bounds are multiplied by 1.7 to define those for $v_p$. The uniform prior pdf for $\rho$ is bounded between 1 $g/cm^3$ and 4 $g/cm^3$. The likelihood function is set to be an uncorrelated Gaussian distribution with standard deviations represented by red error bars in Figure \ref{fig:swi_truevs_data}b, at a level of about 4\% of each datum value which is not uncommon in field studies -- see for example \cite{galetti2017transdimensional} and \cite{fone2024ambient}. The same prior information and likelihood function are used to perform Bayesian inversion under the three different parametrisations, and MH-McMC is employed for each inversion.

\begin{figure}
	\centering\includegraphics[width=\textwidth]{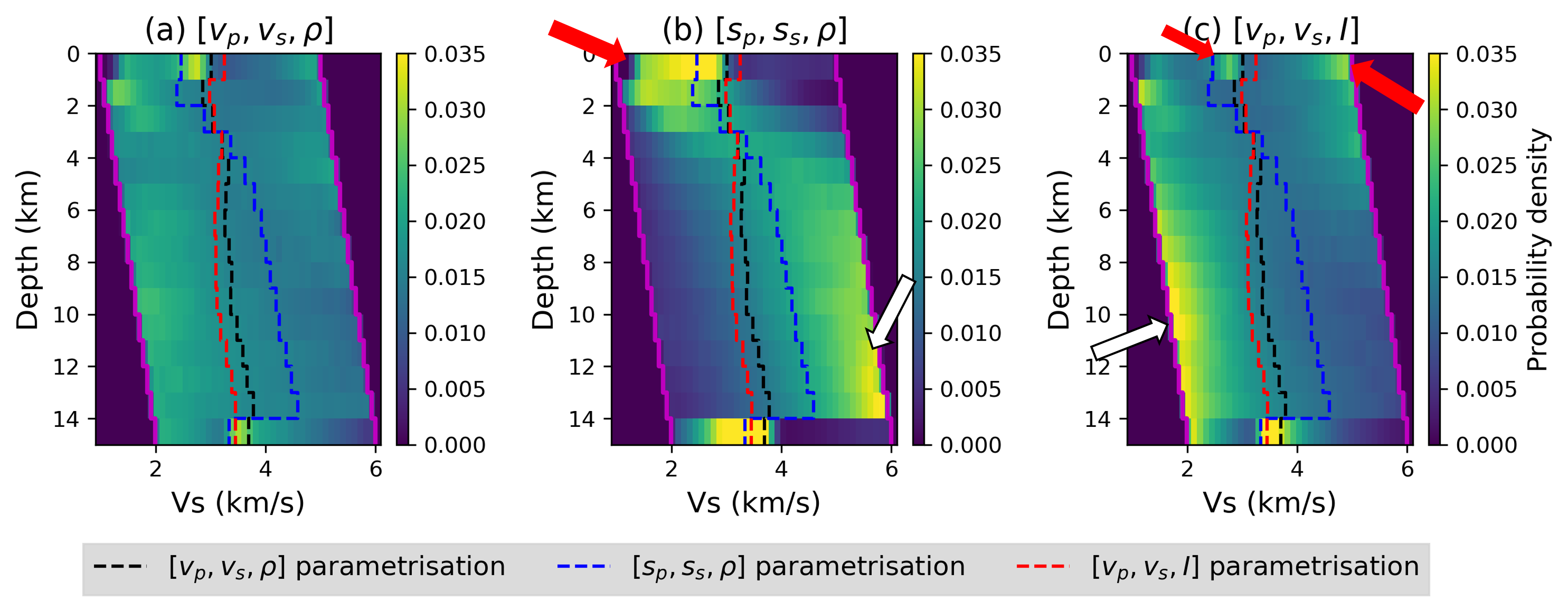}
	\caption{Posterior marginal distributions of shear velocity at each depth, obtained using (a) $[v_p, v_s, \rho]$, (b) $[s_p, s_s, \rho]$, and (c) $[v_p, v_s, I]$ parametrisations. In each panel, dashed lines with different colours display the posterior mean models from different parametrisations.}
	\label{fig:swi_posterior_marginals}
\end{figure}

Figure \ref{fig:swi_posterior_marginals} displays the posterior marginal distributions of the inverted $v_s$ models. As expected, different results are obtained when using different parametrisations due to the BK-inconsistency. At depths beyond 4 km, the posterior marginal pdf's are biased towards high $v_s$ in Figure \ref{fig:swi_posterior_marginals}b, and are biased towards low $v_s$ in Figure \ref{fig:swi_posterior_marginals}c, compared to those in Figure \ref{fig:swi_posterior_marginals}a, as highlighted by two white arrows. This effect is more significant in deeper layers where surface wave dispersion data typically offer less tight constraints than on comparable layer thicknesses in shallower layers, increasing the impact of inconsistent prior pdf's on the posterior solutions. Therefore, the inconsistent results partly reflect the corresponding inconsistency when evaluating the same prior information under the three parametrisations. However, the full explanation has additional subtleties.

In Bayesian inversion, posterior solutions combine information from both the inconsistent prior distributions and the same observed data: the prior distribution is updated to the posterior pdf to provide model solutions that fit the data. While data at only longer periods are significantly sensitive to deeper layers, all data are sensitive to shallow layers. If a deeper velocity is shifted away from the true values in one direction (say, to higher values), the fit to the observed data overall can be improved by shifting shallow velocities in the opposite direction (to lower values). 

Figure \ref{fig:swi_posterior_marginals} shows this trade-off clearly: the posterior marginal pdf's in the near surface (above 3 km) show biases in directions opposite to those at depth, to counteract the effect of the inconsistent prior pdf's from the three parametrisations, so as to still fit the dispersion data. These features are marked by three red arrows: the posterior probability densities in Figure \ref{fig:swi_posterior_marginals}b are higher around the red arrow (at low $v_s$ values) than those in Figure \ref{fig:swi_posterior_marginals}a; the posterior densities are lower around the left red arrow in Figure \ref{fig:swi_posterior_marginals}c (at low $v_s$ values), and are slightly higher around the right red arrow (at high $v_s$ values), compared to those in Figure \ref{fig:swi_posterior_marginals}a. Although the trade-off in Figure \ref{fig:swi_posterior_marginals}c is subtle, it is robust in multiple independent McMC runs. We also plot the posterior mean $v_s$ models obtained from the three parametrisations, represented by dashed black, blue and red lines respectively, in each panel in Figure \ref{fig:swi_posterior_marginals}. Similarly, posterior mean velocity profiles are inconsistent when evaluated under the three parametrisations, and the inconsistencies trade-off between shallower and deeper layers. 

\begin{figure}
	\centering\includegraphics[width=\textwidth]{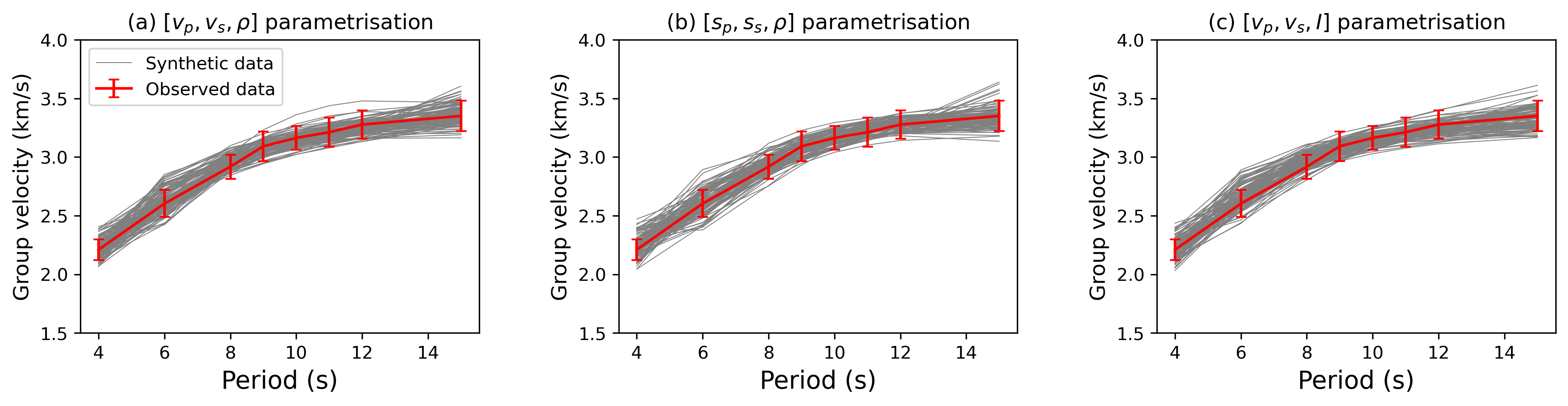}
	\caption{Observed Love wave dispersion curve (red line) and modelled ones (grey lines) obtained using posterior samples obtained from different parametrisations.}
	\label{fig:swi_posterior_synthetic_data}
\end{figure}

We draw 100 posterior samples from each posterior pdf, and calculate the corresponding synthetic dispersion curves, displayed as grey lines in Figure \ref{fig:swi_posterior_synthetic_data}. Despite the inherent inconsistencies among the three sets of posterior solutions, all synthetic curves align almost equally well with the observed dispersion data. This indicates a critical challenge: there is no obvious way to discriminate between the inconsistent posterior solutions in Figure \ref{fig:swi_posterior_marginals} on the basis of the observed dispersion data, which unfortunately are usually the only information available. 

This scenario highlights our significant concern in geophysical inversion. Different parametrisations, for example deployed by different individuals based on personal preferences, can lead to different inversion results, each of which appear to be justifiable based on their fit to observed data. Biased interpretations of subsurface properties will therefore occur, if we ignore the BK-inconsistency problem.

\begin{figure}
	\centering\includegraphics[width=\textwidth]{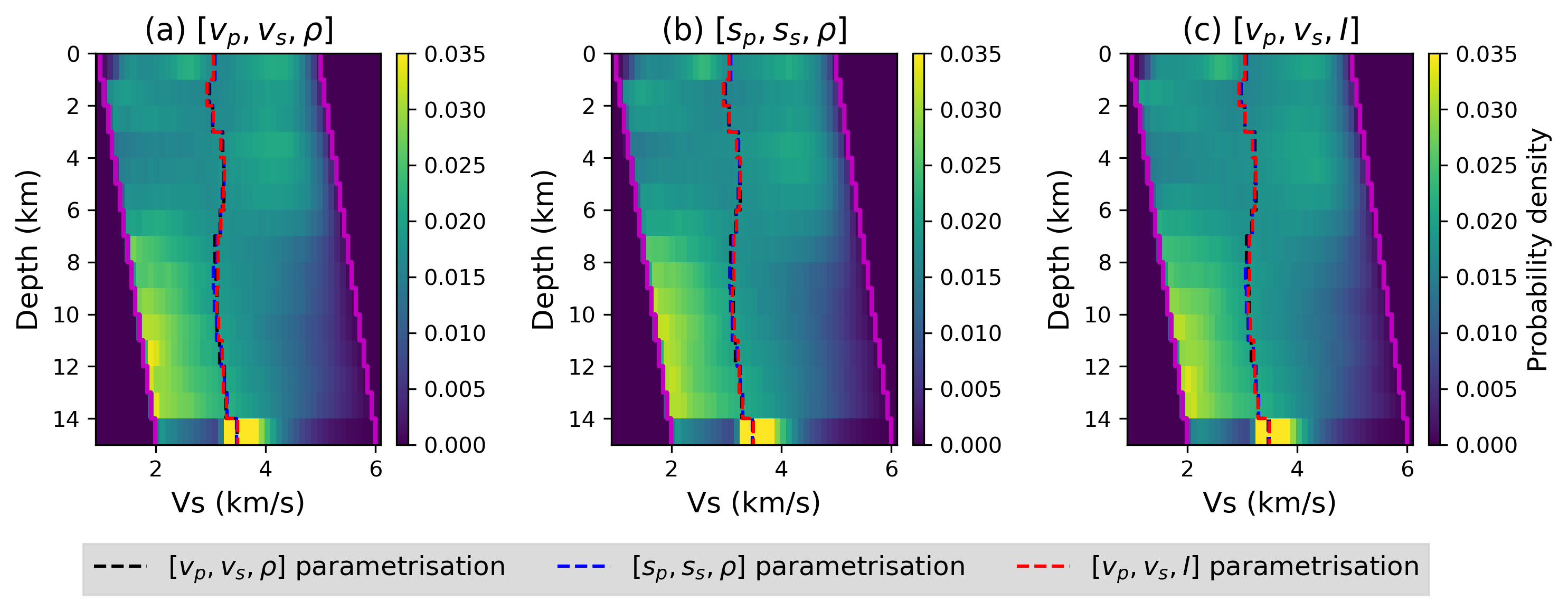}
	\caption{Posterior marginal distributions of shear velocity at each depth, obtained using (a) $[v_p, v_s, \rho]$, (b) $[s_p, s_s, \rho]$, and (c) $[v_p, v_s, I]$ parametrisations, without applying the inter-parameter relations. In each panel, dashed lines with different colours display the posterior mean models from different parametrisations.}
	\label{fig:swi_Grant_posterior_marginals_nocondition}
\end{figure}

To prove that the inconsistent results are caused by the predefined inter-parameter conditions (i.e., $v_p = 1.73v_s$ and $\rho = 0.31(1000v_p)^{0.25}$), we run additional inversion tests without applying these conditions. We introduce additional prior information that $v_p > v_s$ such that any proposed model sample from the Monte Carlo algorithm represents a valid solid Earth structure. Note that this is implicit in the above test since we set $v_p = 1.73v_s$. All other experimental settings are exactly the same as those used to obtain the results in Figure \ref{fig:swi_posterior_marginals}. In Figure \ref{fig:swi_Grant_posterior_marginals_nocondition}, identical posterior solutions are obtained using the three parametrisations. Therefore, Figures \ref{fig:swi_posterior_marginals} and \ref{fig:swi_Grant_posterior_marginals_nocondition} together prove that the BK-inconsistency problem arises due to the introduction of the parameter subspace manifold.

\subsection{Travel time tomography}
\begin{figure}
	\centering\includegraphics[width=0.5\textwidth]{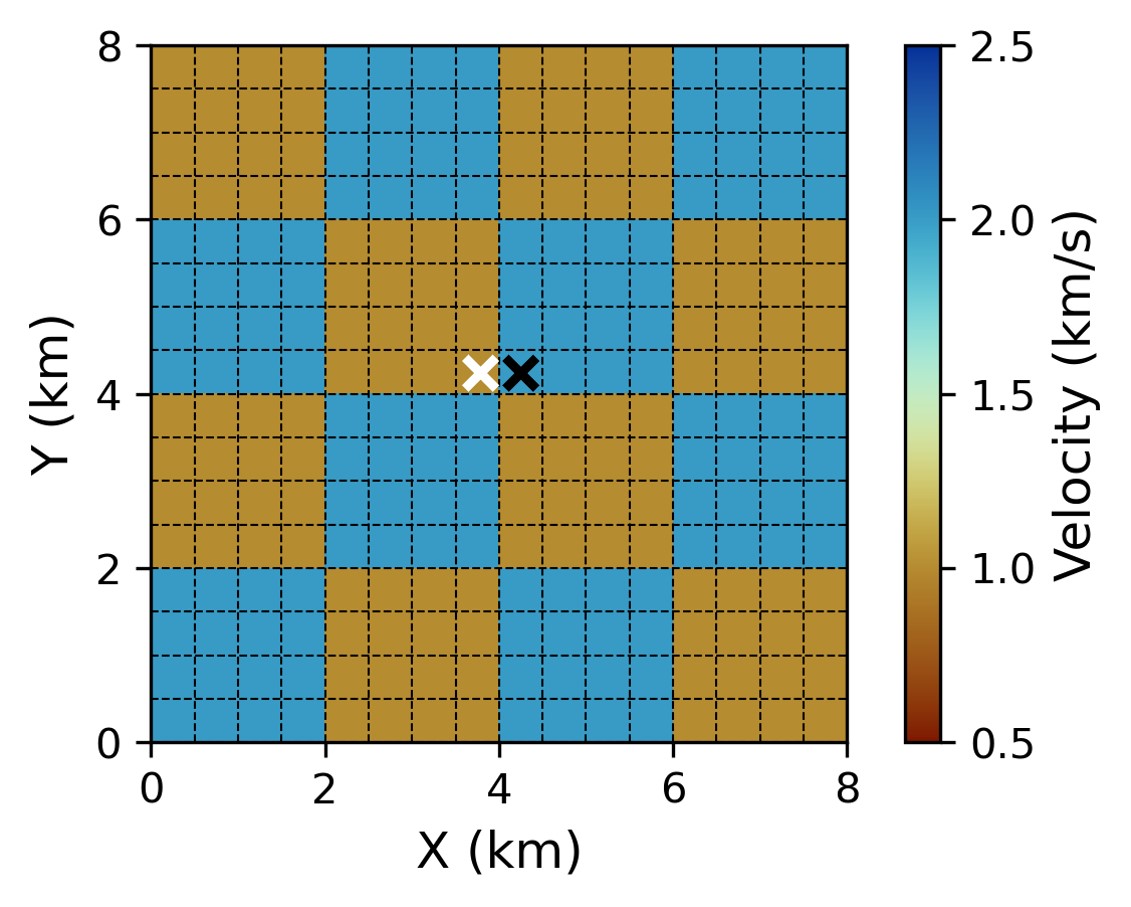}
	\caption{True velocity model used in a travel time tomography example. Low and high velocity values are 1 km/s and 2 km/s, respectively. Dashed black lines show a discretisation of the true velocity model into a grid of 16 $\times$ 16 cells. White and black crosses mark two locations whose posterior marginal distributions are compared in Figures \ref{fig:tomo_marginal_2by2} and \ref{fig:tomo_marginal_4by4}.}
	\label{fig:tomo_true_vel}
\end{figure}

Seismic travel time tomography is a common seismic imaging method that estimates the seismic velocity structure in the interior of a medium such as the Earth's subsurface, using measured first arrival times of seismic waves travelling through the medium between source and receiver locations. We consider a 2D synthetic example which has a chequerboard velocity pattern. The true model is shown in Figure \ref{fig:tomo_true_vel}. We discretise the imaging region into a grid of 16 $\times$ 16 cells with a cell size of 0.5 km in both directions, displayed by dashed black lines. A uniform prior distribution bounded between 0.5 km/s and 2.5 km/s is defined for seismic velocity values in each cell, which encompasses the true velocity values. We use both velocity and slowness parametrisations to represent the prior information. 

\begin{figure}
	\centering\includegraphics[width=\textwidth]{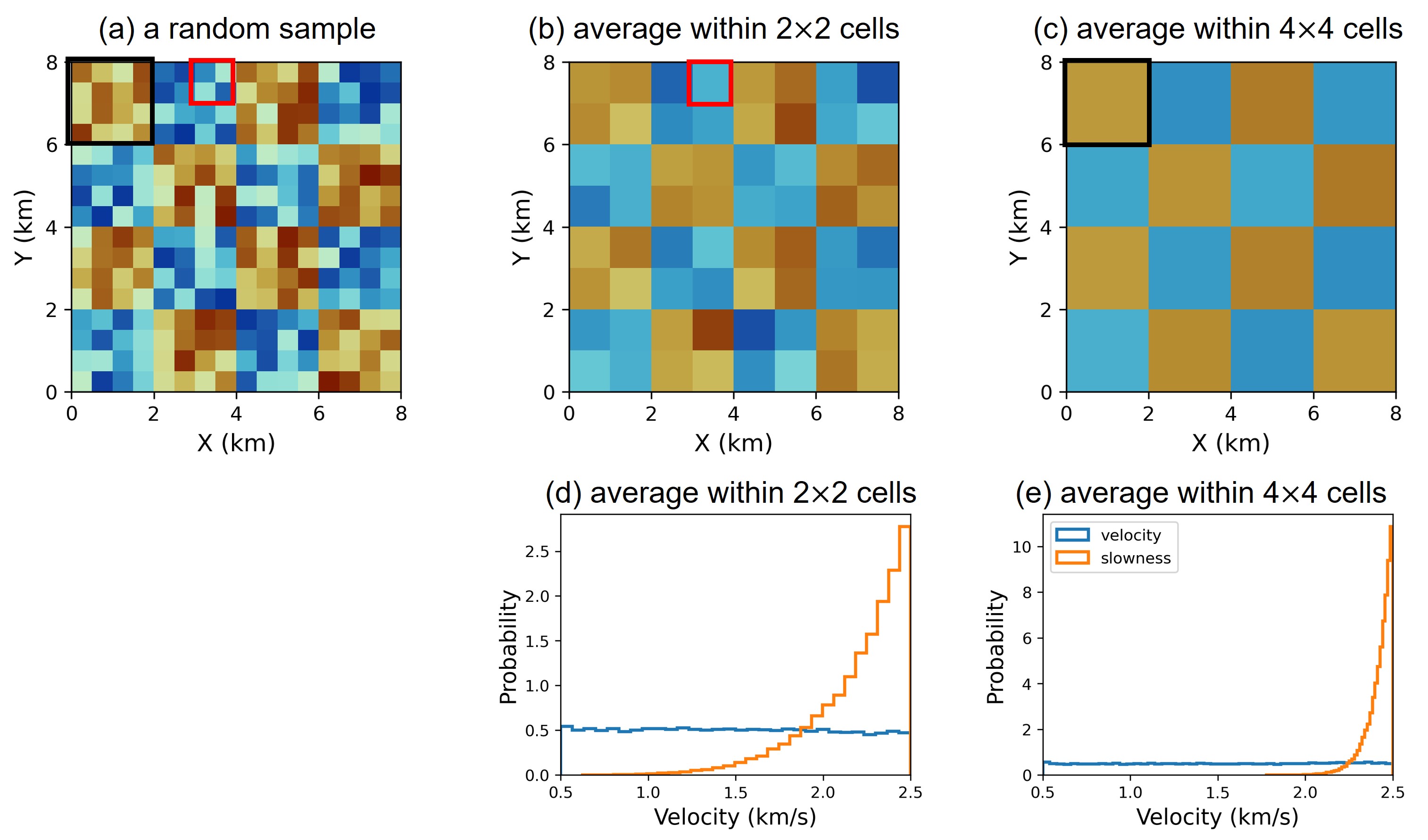}
	\caption{(a) A random model sample from the prior pdf. (b) and (c) Samples obtained by averaging parameter values in (a) within 2 $\times$ 2 or 4 $\times$ 4 cells, respectively, as illustrated by red and black square outlines. (d) and (e) Conditional marginal histograms of velocity values in one arbitrary cell evaluated under velocity and slowness parametrisations.}
	\label{fig:tomo_average_sample}
\end{figure}

We will show that solving a tomography problem using the original grid of 16 $\times$ 16 cells is accurate and provides what we will refer to as high resolution results. Yet it can also be expensive due to the curse of dimensionality \citep{curtis2001prior}. Instead, solving the problem using coarser cells increases the computational efficiency at the expense of accuracy. Techniques such as multilevel McMC take advantages of this situation by performing Monte Carlo sampling over a set of grids with different grid cell sizes. The results are then combined to provide Monte Carlo estimators under the original fine grid system which results in a significantly more efficient sampling algorithm \citep{dodwell2019multilevel}. In this example, we analyse this idea by defining two sets of coarser grids, in which parameter (velocity or slowness) values within groups of 2 $\times$ 2 or 4 $\times$ 4 cells are replaced by their average value, as displayed in Figure \ref{fig:tomo_average_sample}. Figure \ref{fig:tomo_average_sample}a shows a valid prior model sample. Parameter values within 2 $\times$ 2 and 4 $\times$ 4 cell blocks are averaged (such as those outlined by red and black squares in Figure \ref{fig:tomo_average_sample}), and are then used to replace the original parameter values within the respective blocks, as shown in Figures \ref{fig:tomo_average_sample}b and \ref{fig:tomo_average_sample}c. This approach creates two conditional subspace manifolds, in both of which the true chequerboard model can be expressed exactly. The hyper-volume of parameter (sub)space to be explored is thus reduced to save computational cost; however, this also reduces the ability of any proposed Monte Carlo sample to express small scale heterogeneity. Nevertheless, in principle the sampling results can be used to build an unbiased Monte Carlo estimator under the original fine grid using multilevel McMC \citep{dodwell2019multilevel}. 

Since the original model is a valid sample from the prior pdf, so are the new samples since they lie within the prior bounds. Thus, the two spaces of averaged models represent parameter subspaces or manifolds within the original space of models. Also note that the three samples in this Figure can represent either velocity or slowness models, therefore the figures deliberately omit specific units to maintain generality.

Such methods are common in practical geophysical applications. For example, well-logging results typically have higher resolution than seismic inversion models. Parameter values of well-logging results are averaged within a certain window size, and are used as prior information for seismic inversions that use coarser grids. In addition, such strategies are often included explicitly within tomographic inversions through methods such as trans-dimensional sampling \citep{green1995reversible, sambridge2006trans, bodin2009seismic, sambridge2025trans}, which include models defined by different grids within a single inversion. Some neural network architectures used for inversion and inference or simply to emulate functional relationships, are introduced for similar purposes, such as the pooling layers that are used to down-sample feature maps coming from previous layers, thereby generating new feature maps with condensed features \citep{gholamalinezhad2020pooling}. And finally, any finite spatial parametrisation representing properties of a physical medium represents an up-scaled resolution when compared to the true interior structure of the material, since the latter is likely to be heterogeneous even at effectively infinitely small length-scales. Therefore, the averaging process described above and illustrated in Figure \ref{fig:tomo_average_sample} reflects broad practice in tomographic problems.

We sample the uniform prior distribution within the two subspace manifolds defined above, under the velocity and slowness parametrisations, respectively. After sampling, slowness model samples are transformed back to velocity values for comparison by taking their reciprocal values. Figures \ref{fig:tomo_average_sample}d and \ref{fig:tomo_average_sample}e display the marginal histograms of velocity values observed in one cell (which is representative of the behaviour in all other cells). Again, the two parametrisations yield inconsistent prior pdf's given identical prior information and conditional constraints. The inconsistency is more pronounced in Figure \ref{fig:tomo_average_sample}e than \ref{fig:tomo_average_sample}d; this may be attributable to the lower dimensionality of the parameter subspace in the former, which is investigated further below.

To demonstrate the extent to which this inconsistent prior information affects Bayesian tomographic posterior solutions, we perform several inversion tests in which we progressively reduce the number of receivers (which also reduces the number of sources since we imitate an ambient noise tomography study in which receivers are also used as virtual sources). This reduction -- adjusting the receiver number between 16, 12, 8 and 0 -- diminishes the information available from data, leading to increasing posterior inconsistencies caused by the inconsistent prior pdf's. We compare the inversion results obtained by averaging parameter values within 2 $\times$ 2 and 4 $\times$ 4 cells. For each case we define a Gaussian likelihood function but do not add Gaussian random noise to observed data. This ensures that the observed data can be matched perfectly by synthetic data generated from a particular prior sample, meaning that the true model has a non-zero posterior probability density value. While this is commonly referred to as an `inverse crime' in examples used to demonstrate the robustness of methods to measurement noise, that is not our goal. Here we show that even in this near-perfect case with no noise at all, the BK-inconsistency has a significant effect. In noisy synthetic studies or when using real data, this would be conflated with noise-related effects, the consequences of which have been studied in many previous papers so need no replication here \cite[e.g., ][]{bodin2012transdimensionaltomo, zunino2023hmclab}.

\begin{figure}
	\centering\includegraphics[width=\textwidth]{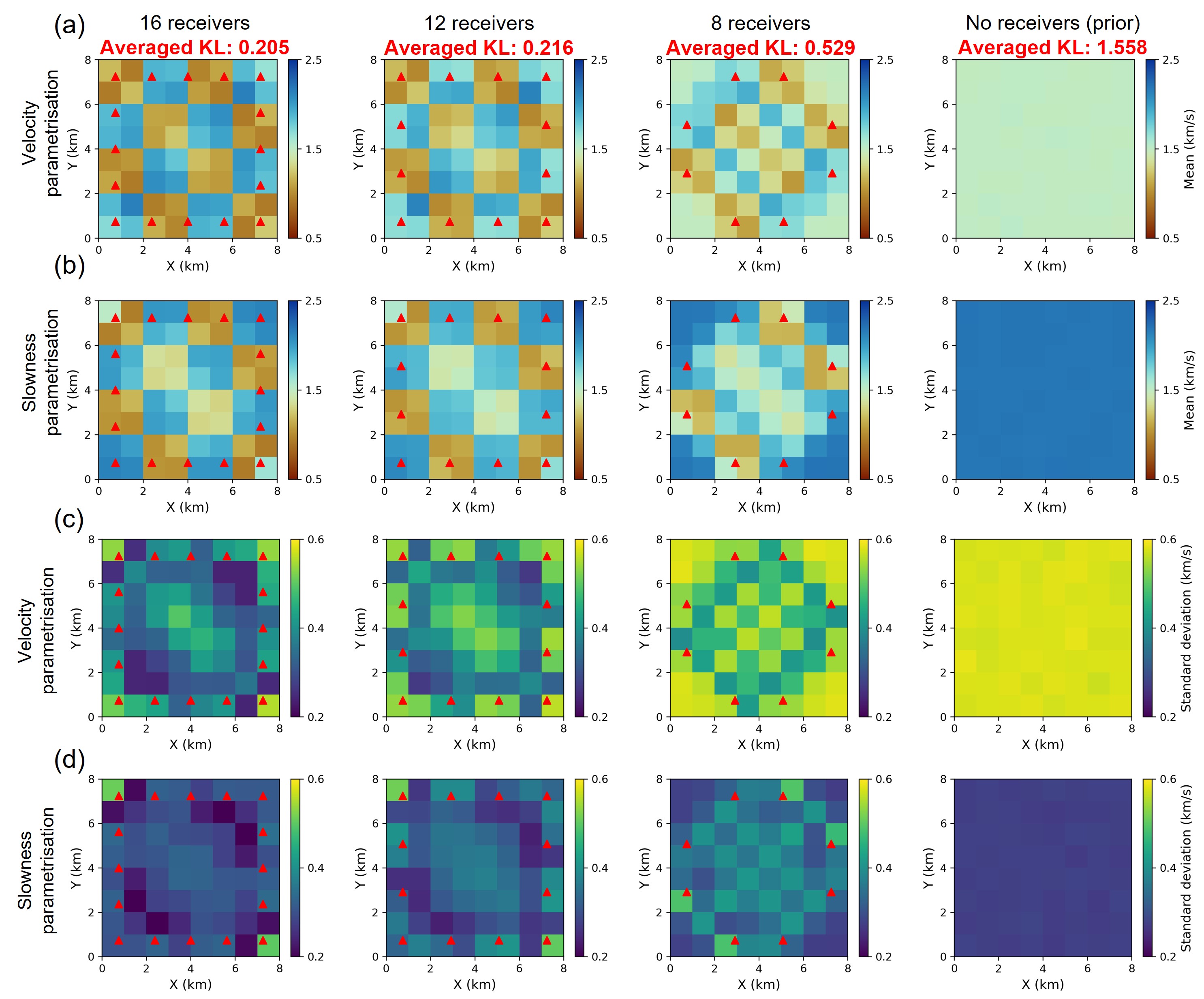}
	\caption{Posterior mean velocity maps given the condition that parameter values within 2 $\times$ 2 cells are averaged, using (a) velocity and (b) slowness parametrisations, respectively. (c) and (d) Posterior standard deviation maps corresponding to the mean velocity maps in (a) and (b). From left to right, each column shows the inversion results obtained using 16, 12 and 8 receivers, as well as no receivers on the right (i.e., results from prior pdf's). Receiver (as well as source) locations are marked by red triangles in each panel. Red text at the top shows the KL-divergence between one-dimensional marginal pdf's in each cell obtained using velocity and slowness parametrisations, averaged across all cells.}
	\label{fig:tomo_mean_std_2by2}
\end{figure}

Figure \ref{fig:tomo_mean_std_2by2} presents the results of averaging parameter values within 2 $\times$ 2 cells. Figures \ref{fig:tomo_mean_std_2by2}a and \ref{fig:tomo_mean_std_2by2}b show the posterior mean velocity maps found by performing inversions using the velocity and slowness parametrisations, respectively, and Figures \ref{fig:tomo_mean_std_2by2}c and \ref{fig:tomo_mean_std_2by2}d display the corresponding standard deviation maps. From left to right, each column represents the results obtained using 16, 12 and 8 receivers, as well as no receivers (the results from the prior pdf's), respectively. Red triangles mark the locations of the receivers (as well as sources) for each case; the data set consists of travel times of first-arriving waves travelling between all source-receiver pairs. Inconsistencies between the results from different parametrisations become more obvious as the number of receivers decreases, because the reduction in data information increases the relative dominance of the inconsistent prior distributions on the posterior solutions. 

\begin{figure}
	\centering\includegraphics[width=\textwidth]{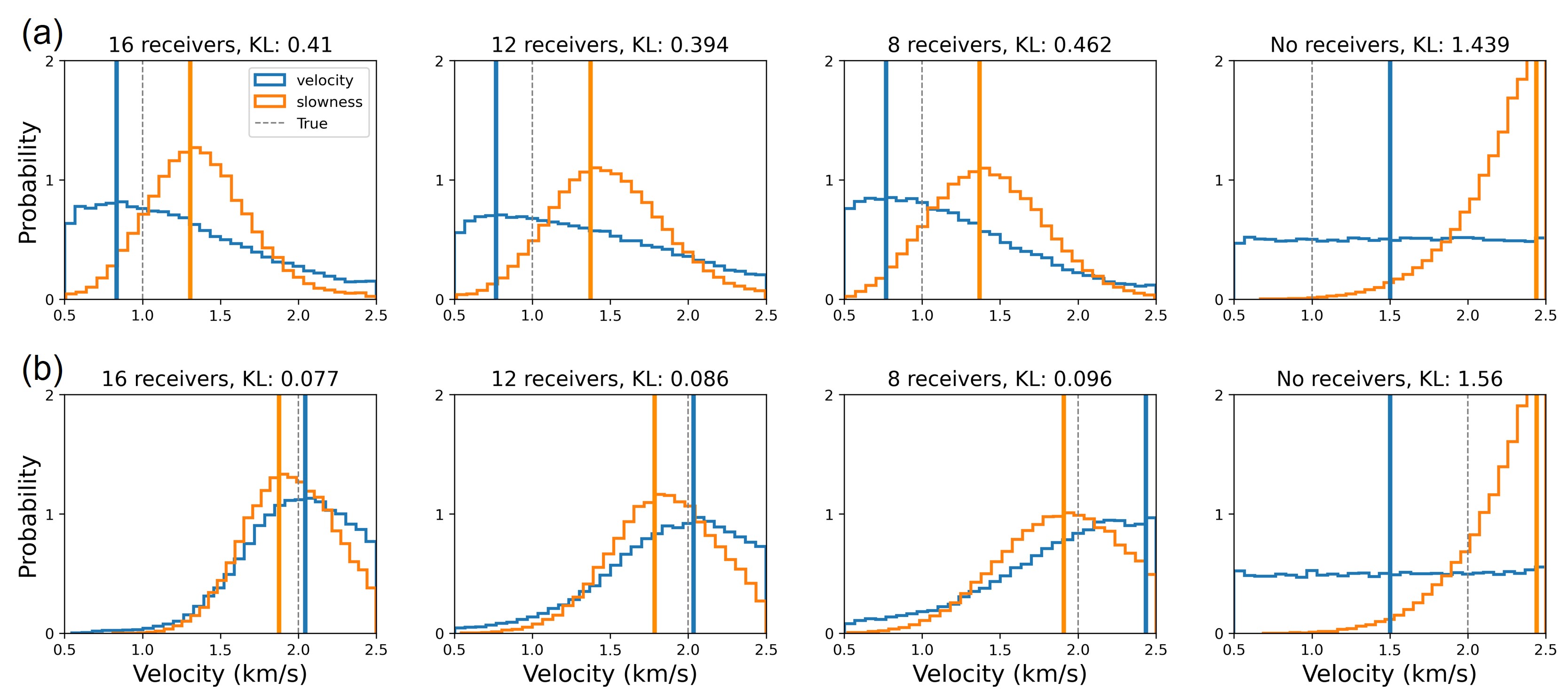}
	\caption{Posterior marginal pdf's at locations marked by (a) white cross and (b) black cross in Figure \ref{fig:tomo_true_vel}. The results in each column correspond to the same column in Figure \ref{fig:tomo_mean_std_2by2}. In each panel, blue and orange histograms show marginal pdf's from velocity and slowness parametrisations, respectively, and blue and orange vertical lines highlight differences in the corresponding \textit{maximum a posteriori} solutions. Dash grey line represents the true velocity value. In each panel, the KL-divergence between the two histograms is provided in the title.}
	\label{fig:tomo_marginal_2by2}
\end{figure}

In Figures \ref{fig:tomo_marginal_2by2}a and \ref{fig:tomo_marginal_2by2}b we compare the posterior marginal pdf's at two representative locations denoted by a white cross and a black cross in Figure \ref{fig:tomo_true_vel}, respectively. Blue and orange histograms approximate posterior marginal pdf's obtained from the velocity and slowness parametrisations, and the corresponding vertical lines highlight differences in the maximum \textit{a posteriori} (MAP) solutions for each case. Vertical dashed grey lines show the true velocity values. In each panel, we calculate the Kullback-Leibler (KL) divergence \citep{kullback1951information} between the two presented marginal pdf's, and the value is provided in panel titles. KL-divergence is a non-negative value that is often used to quantify the difference (distance) between two probability distributions -- a larger value indicates a more significant inconsistency between the pdf's. In addition, we calculate the averaged KL-divergence across all cells, as indicated by the red text at the top of Figure \ref{fig:tomo_mean_std_2by2}. All of the features in Figures \ref{fig:tomo_mean_std_2by2} and \ref{fig:tomo_marginal_2by2} prove that the degree of posterior inconsistency intensifies with the reduction in the number of receivers (thus also the travel time data available). This effect can also be seen most keenly in parts of the model with lowest resolution, such as the corners of the model in the third column (8 receivers). This indicates that even if the data constrain one part of a model well, the inconsistency will prevail in less well resolved areas.

\begin{figure}
	\centering\includegraphics[width=\textwidth]{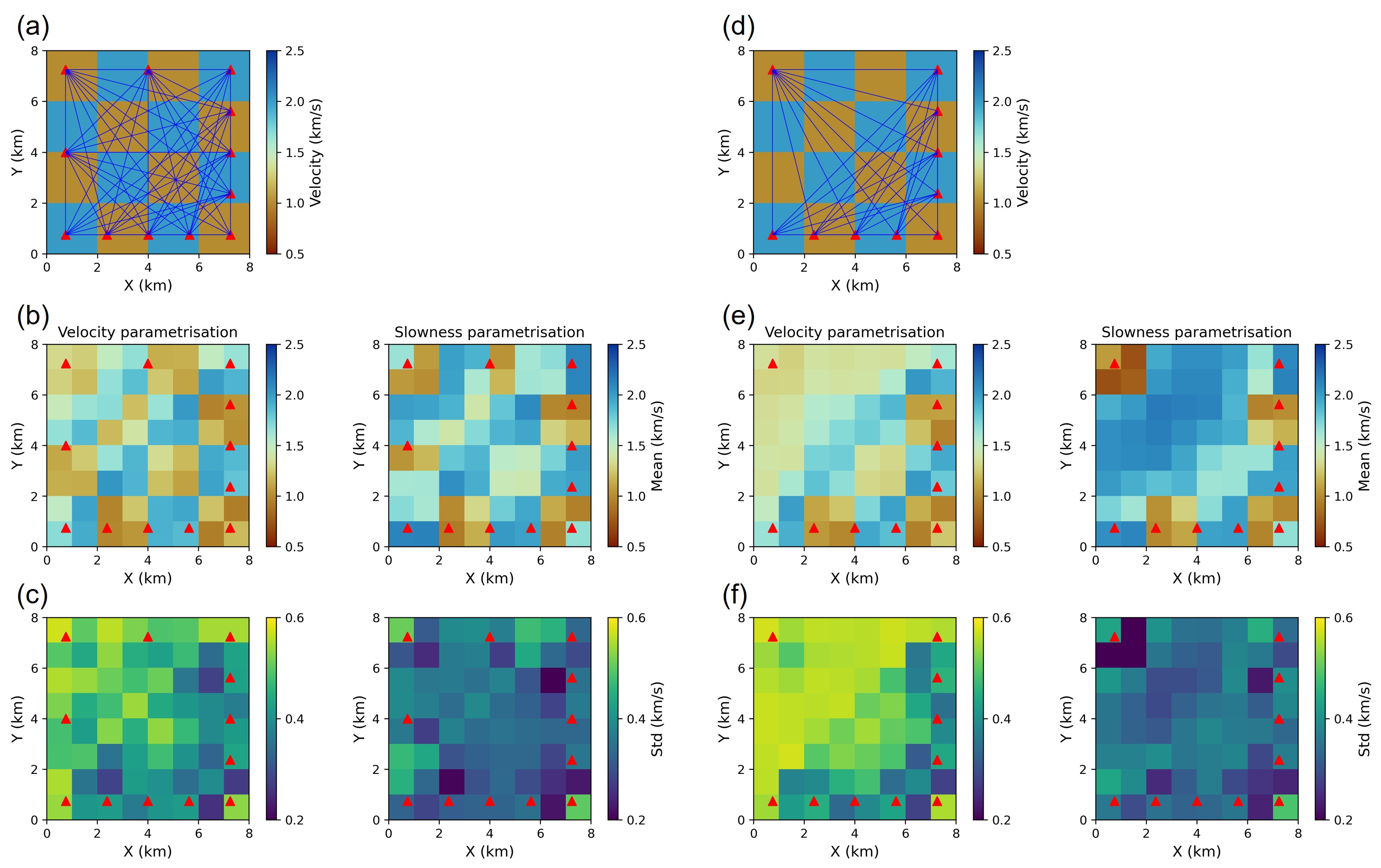}
	\caption{(a) and (d) Two uneven station configurations, with blue lines showing straight ray paths considered for each inversion. (b) and (e) Posterior mean maps obtained using each parametrisation. (c) and (f) Corresponding posterior standard deviation maps.}
	\label{fig:tomo_mean_std_2by2_halfmodel_coverage}
\end{figure}

To prove the latter statement we perform two additional tests with uneven receiver (virtual source) configurations, as shown by red triangles in Figures \ref{fig:tomo_mean_std_2by2_halfmodel_coverage}a and \ref{fig:tomo_mean_std_2by2_halfmodel_coverage}d. Blue lines that connect each pair of receivers represent straight ray paths (an approximation to the true density of ray coverage) considered in each case. Note that actual ray paths for each proposed sample are obtained by tracing the arrival time field from a receiver back to a virtual source and can thus be curved. For both cases, the bottom-right part of the true model is covered by dense rays, and is thus well resolved by travel time data; the top-left part of the model is interrogated by fewer rays. Figures \ref{fig:tomo_mean_std_2by2_halfmodel_coverage}b and \ref{fig:tomo_mean_std_2by2_halfmodel_coverage}e show the posterior mean maps obtained using the two parametrisations, respectively, and Figures \ref{fig:tomo_mean_std_2by2_halfmodel_coverage}c and \ref{fig:tomo_mean_std_2by2_halfmodel_coverage}f show the corresponding standard deviation maps. Again, inconsistent probabilistic results are obtained for each case when solving the inversion under different parametrisations, and these results clearly show that the inconsistency dominates in regions with lower information imposed by the data, more than it affects regions with good data constraints. This is important because acquisition geometries in real problems are always uneven, so this effect is very likely to affect interpretation of tomographic results evaluated under different parametrisations.

\begin{figure}
\centering\includegraphics[width=\textwidth]{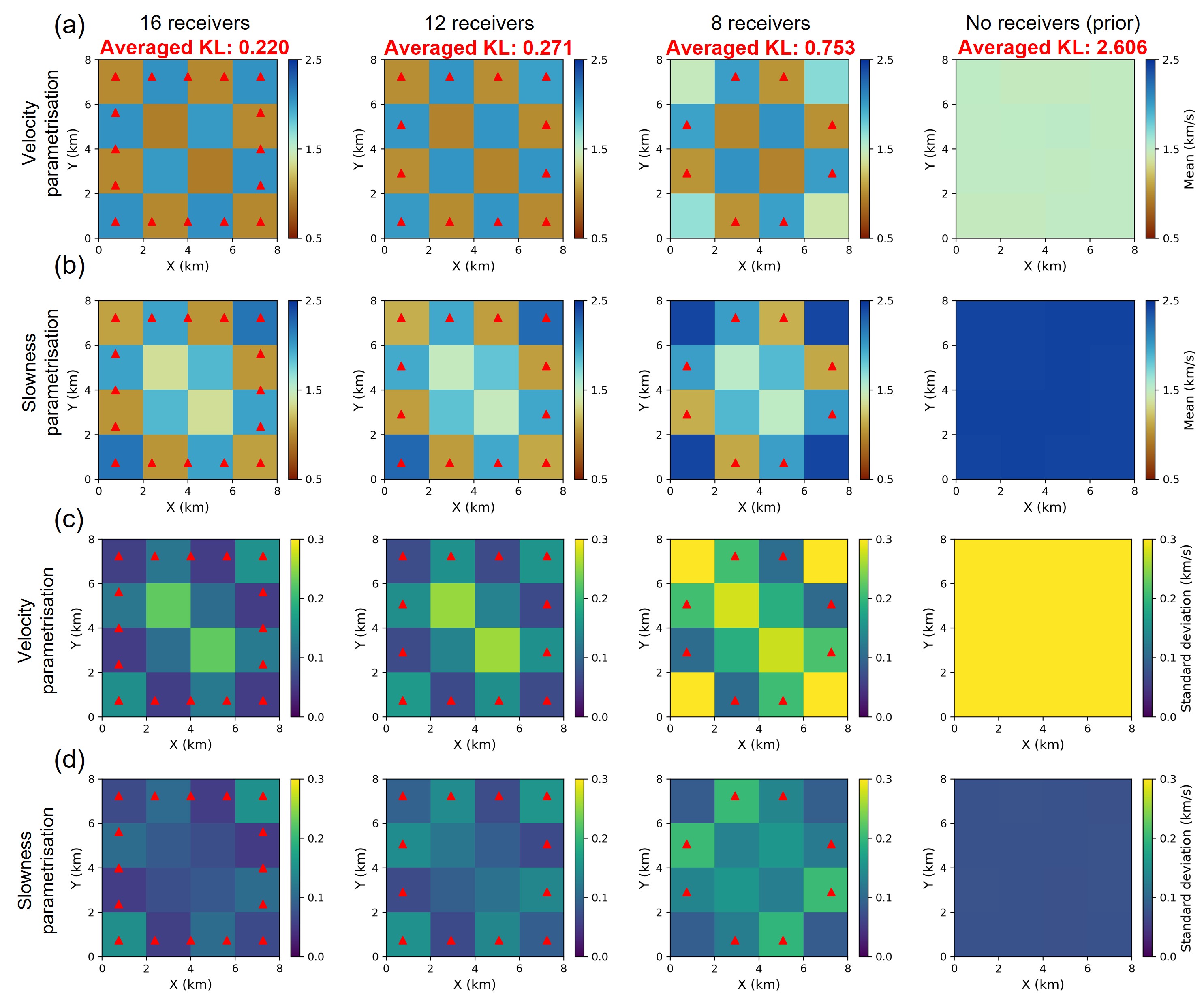}
\caption{Posterior statistics given the condition that parameter values within 4 $\times$ 4 cells are averaged. Key as in Figure \ref{fig:tomo_mean_std_2by2}.}
\label{fig:tomo_mean_std_4by4}
\end{figure}

\begin{figure}
\centering\includegraphics[width=\textwidth]{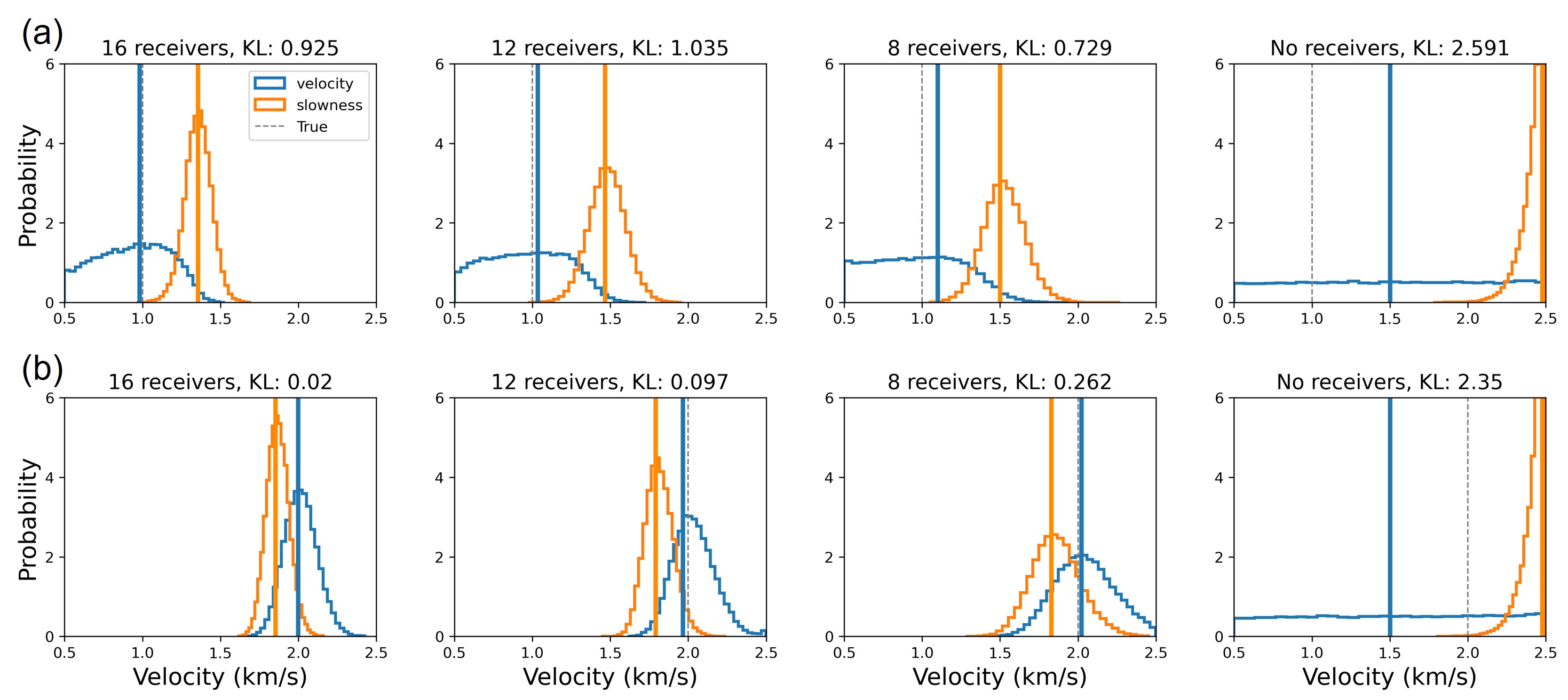}
\caption{Posterior marginal pdf's given the condition that parameter values within 4 $\times$ 4 cells are averaged. Key as in Figure \ref{fig:tomo_marginal_2by2}.}
\label{fig:tomo_marginal_4by4}
\end{figure}

Figures \ref{fig:tomo_mean_std_4by4} and \ref{fig:tomo_marginal_4by4} display similar results as those in Figures \ref{fig:tomo_mean_std_2by2} and \ref{fig:tomo_marginal_2by2}, but with parameter values averaged within 4 $\times$ 4 cell blocks. Note that the overall standard deviation values in this case are lower than those obtained by averaging parameter values within 2 $\times$ 2 cells. This reflects a common compromise that exists between model resolution and model uncertainty in imaging inverse problems \citep{backus1967numerical, backus1968resolving}. Nevertheless, we draw similar conclusions to those obtained above from Figures \ref{fig:tomo_mean_std_2by2} and \ref{fig:tomo_marginal_2by2}. 

An interesting observation in Figures \ref{fig:tomo_marginal_2by2} and \ref{fig:tomo_marginal_4by4} is that in each column (except for the rightmost column with no data), the two sets of posterior marginal pdf's obtained using the slowness parametrisation (orange histograms) exhibit biases in opposite directions in the top and bottom panels, respectively, when compared to those using the velocity parametrisation (blue histograms). This is because the introduction of observed data prompts a trade-off in parameter values between adjacent cells. This trade-off acts to `cancel out' the overall impact of the BK-inconsistency, allowing posterior samples from both parametrisations to fit the data; this trade-off is the travel-time tomographic analogue to the surface-wave dispersion inversion results presented in Figure \ref{fig:swi_posterior_marginals}.

Finally, if we compare the results from averaging values in 2 $\times$ 2 cells to those from averaging values in 4 $\times$ 4 cells, especially reflected in the averaged KL-divergence values, we observe an increase in the inconsistencies of the posterior pdf's in the latter case. Since averaging parameter values in 4 $\times$ 4 cells can be interpreted as averaging values in 2 $\times$ 2 cells twice (one over a smaller cell size and another over a larger cell size), we can understand intuitively that the inter-parameter conditions imposed by averaging values in 4 $\times$ 4 cells are more stringent than that imposed by averaging in 2 $\times$ 2 cells. Particularly, given that no noise has been added to the observed data and that the true model is a valid solution under all parametrisation schemes considered in this example, the posterior solutions should ideally encompass the true model value. However, as depicted in the right-hand panels of Figure \ref{fig:tomo_marginal_4by4}, the true model value is almost excluded from the posterior pdf's using the slowness parametrisation. This is purely due to the BK-inconsistency. Therefore, this example demonstrates that more substantial inconsistencies arise in posterior solutions where either the information derived from data is reduced or stronger inter-parameter conditions are imposed. 

Perhaps one of the most important results exhibited in these examples is the large change in MAP solutions that are obtained using different parametrisations in models (or parts of models) that are relatively poorly constrained by data. Tomographers often use gradient descent algorithms to find MAP models, rather than searching for full Bayesian estimators such as means. This is because they wish to find models that fit the data, and there is no guarantee that mean models so do. When problems are assumed to be linear with Gaussian data and prior uncertainties, the MAP and mean models coincide \citep{tarantola2005inverse}. However, travel time tomographic problems are often highly nonlinear, and the prior pdf's herein are far from Gaussian; even if a pdf was Gaussian under one parametrisation, it would not be Gaussian under another \citep{mosegaard2016inverse}. As a result, strong biases exist in one MAP solution compared to another found using exactly the same information but different parameters. Given that both parametrisations used in the above examples are employed commonly by tomographers (in addition to a variety of others such as wavelet or Fourier bases, in Cartesian or spherical geometries), it is certainly possible that the BK-inconsistency contributes to the inconsistency in results observed from different groups of practitioners \cite[e.g.,][]{fichtner2025high}.

\section{Discussion}
The BK-inconsistency arises when attempting to define the probabilities of any model on a lower-dimensional conditional subspace as the limit of the probability density values in full parameter space at a sequence of points that converges towards the model \citep{kolmogorov1933foundations}. There are infinitely many such sequences of points which converge towards the same model along different paths, as illustrated schematically by the white arrows in Figure \ref{fig:2d_v1eq2v2}a, and these produce different probability values in the limit. 

\begin{figure}
	\centering\includegraphics[width=\textwidth]{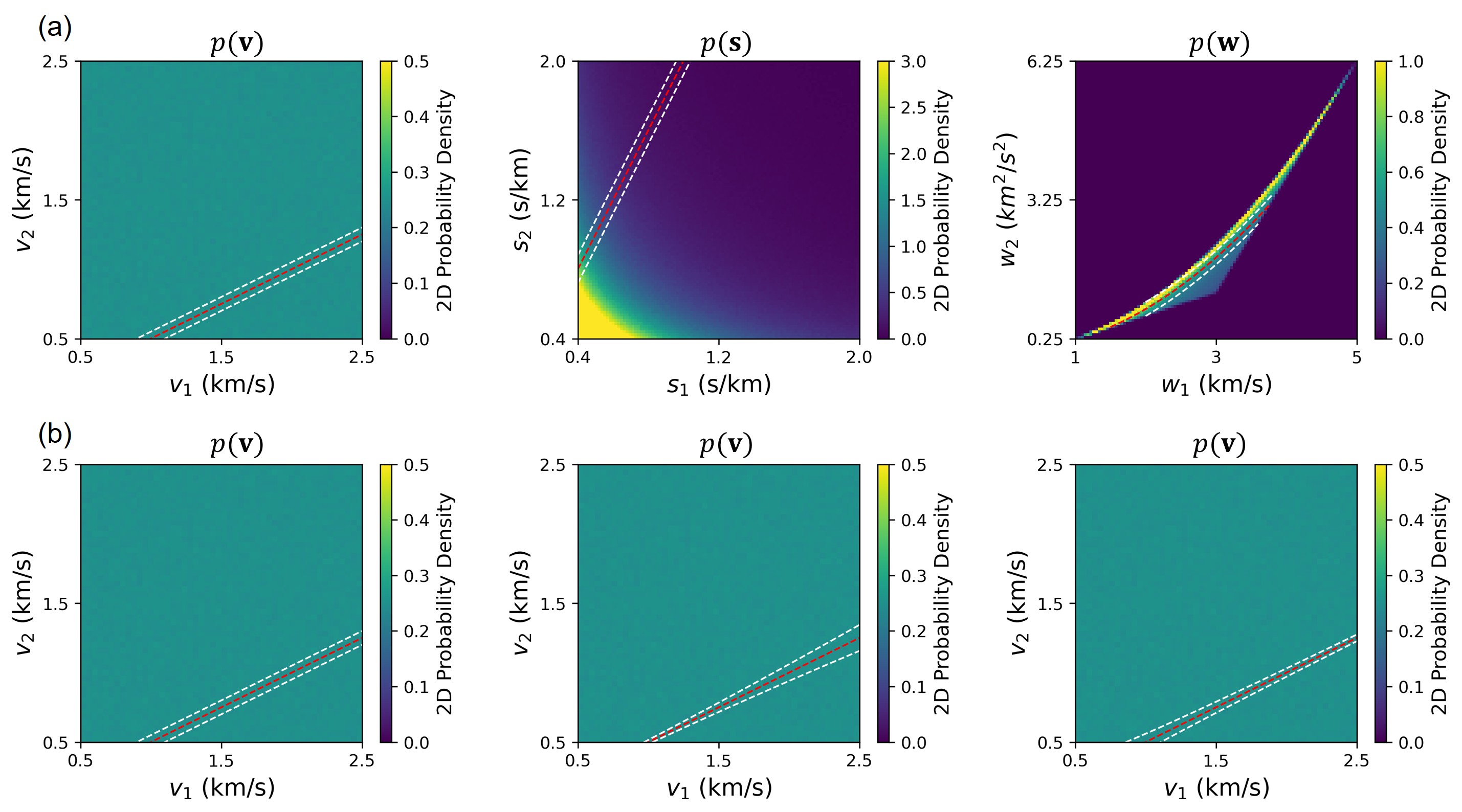}
	\caption{(a) 2D pdf's corresponding to those in Figures \ref{fig:2d_v1eq2v2}a, \ref{fig:2d_v1eq2v2}b and \ref{fig:2d_v1eq2v2}c. Dashed white lines in each panel show two bounds of each interval represented in equation \ref{eq:approach_vsw}. (b) Corresponding results after transforming the pdf's back to the velocity parametrisation. Dashed white lines show the two bounds in equation \ref{eq:approach_v}.}
	\label{fig:2d_v1eq2v2_approach}
\end{figure}

\begin{figure}
	\centering\includegraphics[width=\textwidth]{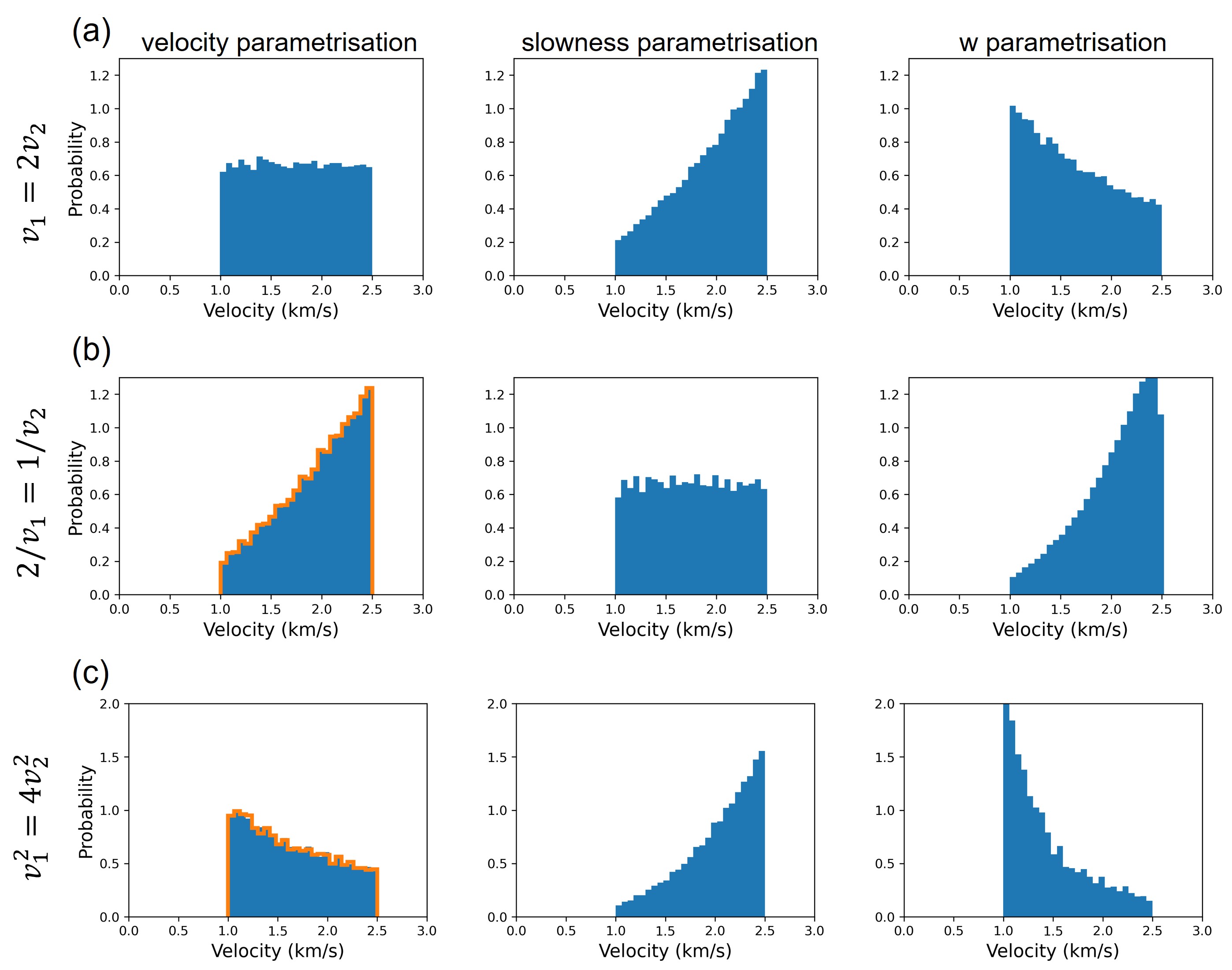}
	\caption{Condition marginal pdf's of parameter $v_1$ given the condition $v_1=2v_2$ evaluated from three different parametrisations. (a), (b) and (c) represent the results evaluated by approaching the manifold subspace linearly through $v_1=2v_2$, reciprocally through $2/v_1 = 1/v_2$ and quadratically through $v_1^2 = 4v_2^2$. Note that panel (a) and Figure \ref{fig:2d_3conditions_1d_marginals}a display the same results. For better comparison, orange histograms in left panels in (b) and (c) correspond to the two blue histograms in middle and right panels in (a).}
	\label{fig:2d_v1eq2v2_different_way_approach_marginals}
\end{figure}

In Figure \ref{fig:2d_v1eq2v2}, we demonstrated that evaluating conditional probabilities under different parametrisations causes the BK-inconsistency. Here we provide more detail about that 2D toy example. For each parametrisation, the conditional distribution is obtained by sampling the 2D space densely and then selecting samples that satisfy
\begin{equation}
	|v_1-2v_2|<\epsilon_1 \quad, \quad |2s_1-s_2|<\epsilon_2 \quad, \quad |2w_1^2-9w_2|<\epsilon_3
	\label{eq:linearly}
\end{equation}
respectively in the corresponding 2D full space, where $\epsilon_1$, $\epsilon_2$ and $\epsilon_3$ are small positive numbers. Equivalently
\begin{equation}
	\begin{split}
	\dfrac{v_1 - \epsilon_1}{2} &< v_2 < \dfrac{v_1 + \epsilon_1}{2} \\
	2s_1 - \epsilon_2 &< s_2 < 2s_1 + \epsilon_2 \\
	\dfrac{2w_1^2 - \epsilon_3}{9} &< w_2 < \dfrac{2w_1^2 + \epsilon_3}{9}
	\end{split}
	\label{eq:approach_vsw}
\end{equation}
Dashed white lines in Figure \ref{fig:2d_v1eq2v2_approach}a illustrate these intervals with $\epsilon_1=\epsilon_2=\epsilon_3 =0.03$, under three different parametrisations. The width of the three intervals in equation \ref{eq:approach_vsw} is constant for any given values of $v_2$, $s_2$ and $w_2$, respectively. Therefore, we observe that the dashed white lines are parallel to the subspace manifold represented by the dashed red line. Samples within these intervals approach the parameter subspace manifold linearly along each coordinate system. If we transform the slowness and $\mathbf{w}$ parametrisations in equation \ref{eq:approach_vsw} back to the velocity parametrisation
\begin{equation}
	\begin{split}
		\dfrac{v_1 - \epsilon_1}{2} & < v_2 < \dfrac{v_1 + \epsilon_1}{2} \\
		\dfrac{v_1}{2+v_1\epsilon_2} & < v_2 < \dfrac{v_1}{2-v_1\epsilon_2} \\
		\dfrac{2(1+v_2 / v_1)^2 - \epsilon_3 / v_1}{9} & < v_2 < \dfrac{2(1+v_2 / v_1)^2 + \epsilon_3 / v_1}{9}
	\end{split}
	\label{eq:approach_v}
\end{equation}
the width of the interval in the first line is still constant with respect to the value of $v_1$ (or $v_2$). However, the width of the intervals in the second and third lines is not constant: it increases in the second line and decreases in the third line, with increasing $v_1$. Figure \ref{fig:2d_v1eq2v2_approach}b illustrates these intervals under the velocity parametrisation. Since $p(\mathbf{v})$ is defined as a uniform distribution, the area between the two dashed white lines in each case represents the 1D conditional pdf given the condition $v_1=2v_2$, as $\epsilon_{1,2,3} \rightarrow 0$, resulting in the conditional pdf's displayed in Figures \ref{fig:2d_v1eq2v2}g, \ref{fig:2d_v1eq2v2}h and \ref{fig:2d_v1eq2v2}i. Figure \ref{fig:2d_v1eq2v2_different_way_approach_marginals}a shows the 1D marginal pdf's corresponding to those in Figure \ref{fig:2d_v1eq2v2_approach}b, conditioned on $v_1 = 2v_2$ (note that Figures \ref{fig:2d_v1eq2v2_different_way_approach_marginals}a and \ref{fig:2d_3conditions_1d_marginals}a present the same results).
 
In addition to approaching the subspace manifold linearly along coordinate system as displayed in Figure \ref{fig:2d_v1eq2v2_approach}a and equation \ref{eq:linearly}, we consider approaching the manifold reciprocally (compared to equation \ref{eq:linearly}) through
\begin{equation}
	|2/v_1-1/v_2|<\epsilon_1' \quad, \quad |1/s_1-2/s_2|<\epsilon_2' \quad, \quad |9/w_1^2-2/w_2|<\epsilon_3'
	\label{eq:reciprocally}
\end{equation}
and quadratically through
\begin{equation}
	|v_1^2-4v_2^2|<\epsilon_1'' \quad, \quad |4s_1^2-s_2^2|<\epsilon_2'' \quad, \quad |4w_1^4-81w_2^2|<\epsilon_3''
	\label{eq:quadratically}
\end{equation}
Obviously, equations \ref{eq:linearly}, \ref{eq:reciprocally} and \ref{eq:quadratically} define different directions used to approach exactly the same manifold ($v_1=2v_2$). Similar inequalities to those in equations \ref{eq:approach_vsw} and \ref{eq:approach_v} can be derived for equations \ref{eq:reciprocally} and \ref{eq:quadratically}. Figures \ref{fig:2d_v1eq2v2_different_way_approach_marginals}b and \ref{fig:2d_v1eq2v2_different_way_approach_marginals}c show the obtained marginal pdf's on $v_1$, by approaching the manifold reciprocally and quadratically.

Each column in Figure \ref{fig:2d_v1eq2v2_different_way_approach_marginals} illustrates our explanation of the BK-inconsistency, in which different directions used to approach a subspace manifold yield different conditional pdf's. On the other hand, each row in Figure \ref{fig:2d_v1eq2v2_different_way_approach_marginals} demonstrates another explanation of the BK-inconsistency, in which inconsistency appears when probability densities within parameter subspaces are evaluated under different parametrisations. Fundamentally, these two explanations are equivalent. To explain, if we approach the manifold reciprocally along $2/v_1 = 1/v_2$ (left panel in Figure \ref{fig:2d_v1eq2v2_different_way_approach_marginals}b) under the velocity parametrisation, we have
\begin{equation}
	|\dfrac{2}{v_1}-\dfrac{1}{v_2}|<\epsilon_1' \quad \Leftrightarrow \quad \dfrac{2}{v_1} - \epsilon_1' < \dfrac{1}{v_2} < \dfrac{2}{v_1} + \epsilon_1' \quad \Leftrightarrow \quad \dfrac{v_1}{2+v_1\epsilon_1'} < v_2 < \dfrac{v_1}{2-v_1\epsilon_1'}
	\label{eq:linear_reciprocal}
\end{equation}
The right equation is exactly the same as that in the second line in equation \ref{eq:approach_v}. A similar derivation can be obtained for the case in which we approach the manifold quadratically along $v_1^2 = 4v_2^2$ under the velocity parametrisation. To demonstrate this clearly, the two blue histograms in the middle and right panels in Figure \ref{fig:2d_v1eq2v2_different_way_approach_marginals}a are replotted by orange histograms in the left panels in Figures \ref{fig:2d_v1eq2v2_different_way_approach_marginals}b and \ref{fig:2d_v1eq2v2_different_way_approach_marginals}c, respectively. The corresponding results are almost identical, verifying that the two explanations are equivalent.

\begin{figure}
	\centering\includegraphics[width=\textwidth]{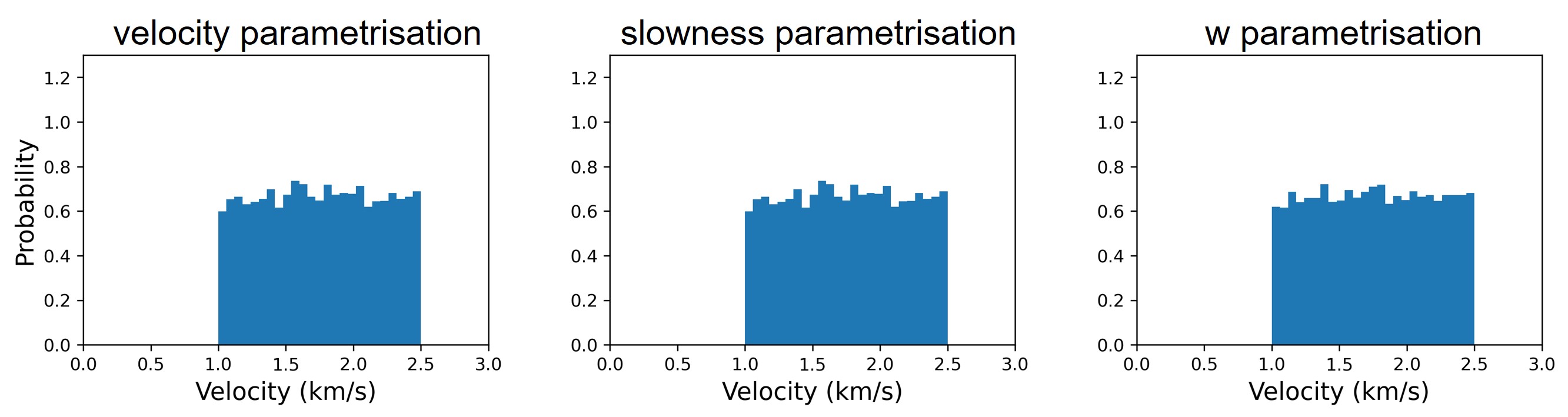}
	\caption{Condition marginal pdf's evaluated by approaching the manifold subspace ($v_1=2v_2$) along particular directions (a) $v_1=2v_2$, (b) $1/s_1=2/s_2$ and (c) $\dfrac{w_1 \pm \sqrt{w_1^2-4w_2}}{2}=w_1 \mp \sqrt{w_1^2-4w_2}$ (equation \ref{eq:inverse_s_w}) that provide consistent probabilistic results.}
	\label{fig:2d_v1eq2v2_approach_inverse_transform}
\end{figure}

More generally, consider a random variable with a given parametrisation $\mathbf{m}$, and define a new parametrisation $\mathbf{n}=T(\mathbf{m})$ obtained by a transformation function $T(\cdot)$. Suppose that functions $g(\mathbf{m})=0$ and $g'(\mathbf{n})=0$ define the same inter-parameter condition (thus define the same parameter subspace manifold). From above we know that we would obtain inconsistent results when evaluating $p(\mathbf{m}~|~[g(\mathbf{m})=0])$ and $p(\mathbf{n}~|~[g'(\mathbf{n})=0])$. For the $\mathbf{n}$ parametrisation, if one could find a specific direction that represents the inverse transform of $T(\cdot)$ (i.e., $T^{-1}(\mathbf{n})$) and approach the subspace manifold along that direction (that is along $g[T^{-1}(\mathbf{n})]=0$), the conditional pdf $p(\mathbf{n}~|~[g[T^{-1}(\mathbf{n})]=0])$ is consistent with $p(\mathbf{m}~|~[g(\mathbf{m})=0])$. Figure \ref{fig:2d_v1eq2v2_approach_inverse_transform} illustrates this concept. Given the definition of $\mathbf{s}$ and $\mathbf{w}$ in our toy example, we calculate the following inverse transforms
\begin{equation}
	\begin{split}
		[v_1, v_2] & = \left[\dfrac{1}{s_1}, \dfrac{1}{s_2}\right] \\
		[v_1, v_2] & = \left[\dfrac{w_1 \pm \sqrt{w_1^2-4w_2}}{2}, \dfrac{w_1 \mp \sqrt{w_1^2-4w_2}}{2}\right]
	\end{split}
	\label{eq:inverse_s_w}
\end{equation}
We further evaluate $p(\mathbf{s}~|~[1/s_1=2/s_2])$ and $p(\mathbf{w}~|~\left[\dfrac{w_1 \pm \sqrt{w_1^2-4w_2}}{2}=w_1 \mp \sqrt{w_1^2-4w_2}\right])$, leading to the conditional marginal pdf's displayed in Figure \ref{fig:2d_v1eq2v2_approach_inverse_transform}. Consistent pdf's are obtained from the three parametrisations. This provides a potentially viable approach to find consistency in physical inference problems, but it does not circumvent the BK-inconsistency: there is no obvious reason why we should choose one path over another in any particular parametrisation.

The toy example can also be interpreted as a 2D inverse problem, in which we wish to estimate the unknown model parameter $\mathbf{v} = [v_1, v_2]^T$. The forward problem is defined as $d = v_1 - 2v_2$ which calculates the difference between the first parameter value and twice of the second parameter value. Define a uniform prior distribution on $\mathbf{v}$ and a uniform likelihood function. The posterior distribution given an observed datum of $d_{obs} = 0$, evaluated under three different parametrisations, can be represented in Figure \ref{fig:2d_v1eq2v2}. Indeed, all examples (the 2D toy example and the three geophysical inversions) presented in this paper show that inverse problem solutions change due to changes of parametrisations. 

\begin{figure}
	\centering\includegraphics[width=\textwidth]{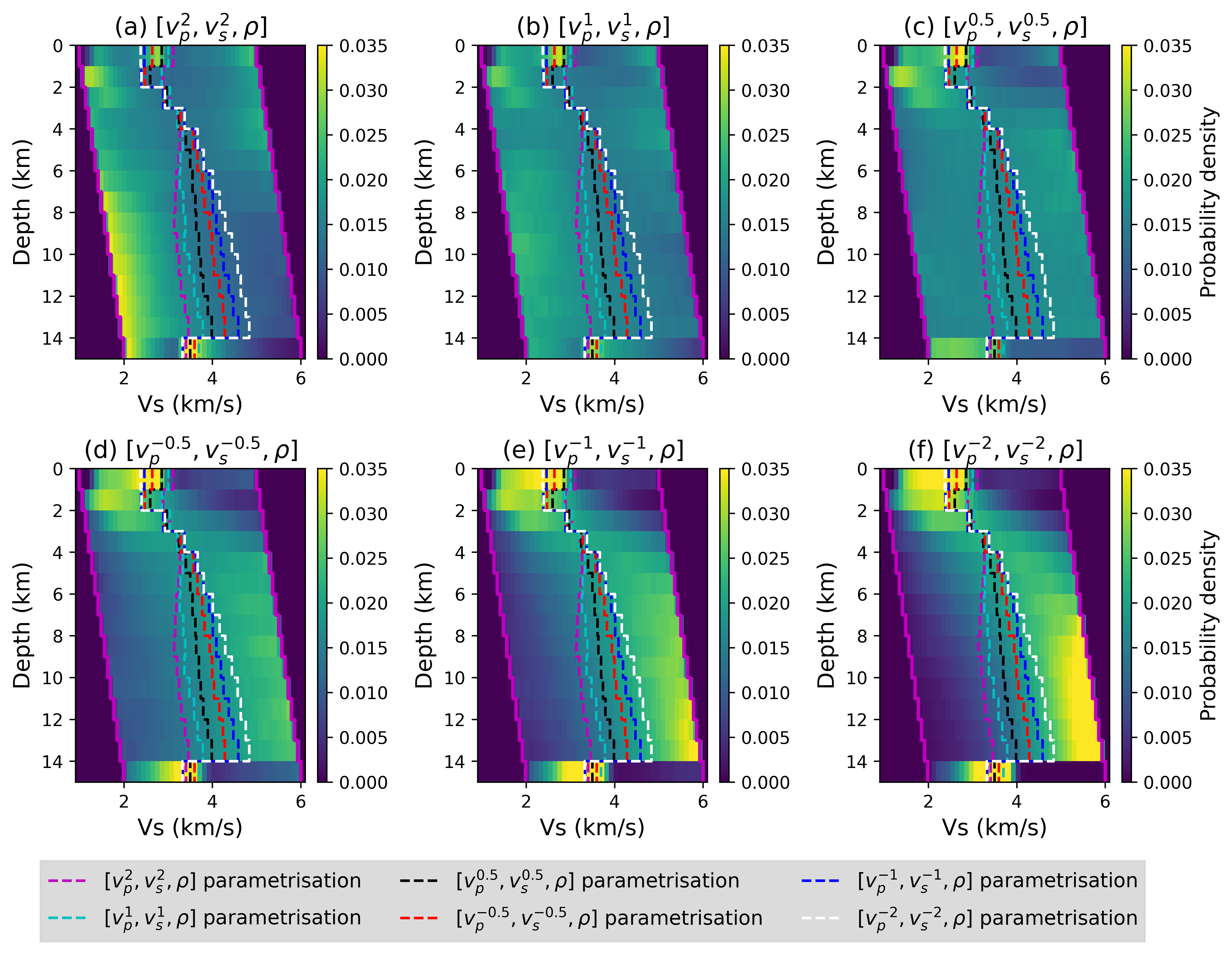}
	\caption{Posterior marginal distributions of shear velocity over depth, obtained using 6 different but equivalent parametrisations. In each panel, identical dashed lines with different colours display the posterior mean models from the six parametrisations.}
	\label{fig:swi_Grant_posterior_marginals_6parametrisations}
\end{figure}

To further demonstrate the impact of our conclusions, we repeat the surface wave inversion example in Section \ref{sec:swi} with the following 6 parametrisations: [$v_p^2, v_s^2, \rho$], [$v_p^1, v_s^1, \rho$], [$v_p^{0.5}, v_s^{0.5}, \rho$], [$v_p^{-0.5}, v_s^{-0.5}, \rho$], [$v_p^{-1}, v_s^{-1}, \rho$], and [$v_p^{-2}, v_s^{-2}, \rho$], respectively -- providing a full set of parametrisations which only vary the power to which the velocity parameters are raised. Posterior distributions of $v_s$ are displayed in Figure \ref{fig:swi_Grant_posterior_marginals_6parametrisations} (note that Figures \ref{fig:swi_Grant_posterior_marginals_6parametrisations}b and \ref{fig:swi_Grant_posterior_marginals_6parametrisations}e are exactly the same as Figures \ref{fig:swi_posterior_marginals}a and \ref{fig:swi_posterior_marginals}b). Clearly, tomographers can \textit{design} completely different but apparently reasonable Bayesian solutions that represent posterior probability densities of the shear velocity structure beneath Edinburgh (55.93$^\circ$N latitude and 3.18$^\circ$W longitude) based on personal preference, simply by changing variables (in this example by changing the power of parameters of interest). This indicates that we could create a range of solutions using different parametrisations, then choose the one which gives our preferred result. Thus we show that existing inverse problem solutions are not robust. 


Trans-dimensional inversion \citep{green1995reversible, sambridge2006trans, bodin2009seismic, bodin2012transdispersion, galetti2018transdimensional, zhang20183} and hierarchical Bayesian inversion \citep{malinverno2004expanded, bodin2012transdimensionaltomo} are commonly used in geophysical inverse problems. These methods typically involve changing parametrisation of either model parameters or data likelihood functions during inversions, so the inversion results generally suffer from the BK-inconsistency problem \citep{mosegaard2024inconsistency}. The travel time tomography example above allows us to understand effects that we might expect to occur when using such methods. For example, comparing solutions found on the manifolds defined by averaging parameter values within 2 $\times$ 2 and 4 $\times$ 4 cells, to that found by solving the tomography problem in the original, full parameter space, represents similar changes in model space geometry and parametrisation to those that occur in trans-dimensional inversion algorithms. Our examples, and \cite{mosegaard2024inconsistency}, show that in the trans-dimensional formulation of this inverse problem, physically inconsistent results are found under different, apparently equivalent model parametrisations. 

Machine learning techniques, such as generative adversarial networks \citep{goodfellow2014generative}, variational autoencoders \citep{kingma2014auto}, and other dimensionality reducing methods \citep{bishop2006pattern}, are widely used to represent a high-dimensional pdf by a lower-dimensional one that contains only important components of parameter variability. Other methods use neural networks to reparametrise unknown properties \citep{ulyanov2018deep, heckel2018deep, wu2025does}. We conjecture that they might suffer from inconsistency problems when applied to physical models.

\cite{mosegaard2024inconsistency} showed that the Bayesian posterior solution is itself a conditional probability distribution of the joint \textit{a priori} distribution of model parameter $\mathbf{m}$ and observed data $\mathbf{d}_{obs}$, conditioned by a forward function $\mathbf{d} = f(\mathbf{m})$. That is to say, the forward function defines a conditional parameter subspace within the joint space, on which Bayesian posterior solutions are sought (evaluated). Therefore, we should in theory observe the BK-inconsistency problem in the whole posterior solution, rather than only on sub-manifolds within that solution space.

However, we have not found a standard geophysical example that clearly demonstrates inconsistent whole posterior pdf's. One reason for this is that in geophysical (or more generally, physical) inference problems, we commonly only define a single analytic or computational forward function based on one specific parametrisation. For example, say we consider both velocity $\mathbf{v}$ and slowness $\mathbf{s}$ parametrisations for a seismic acoustic full waveform inversion problem. The forward function involves predicting seismic waveform data from a model sample by solving the scalar wave equation. For most cases, the only \textit{valid} input for the forward function is a velocity model $\mathbf{v}$. Therefore, for a slowness model sample $\mathbf{s}$, we first convert it into a velocity sample $\mathbf{v}$ through $[v_1, v_2, ..., v_k] = [1/s_1, 1/s_2, .., 1/s_k]$, then substitute it into the forward function. By doing this transformation explicitly, we essentially find an inverse transform and approach the manifold (defined by the forward function) along this particular direction. Based on the above analysis, we will be likely to find consistent results, and thus not observe the BK-inconsistency problem. Nevertheless, if alternative forward functions exist that accept different parametrisations of model parameters and produce synthetic data directly without necessitating inverse transformations between parametrisations beforehand, it is plausible that inconsistent posterior solutions would emerge.

All examples presented in this work primarily address the BK-inconsistency within the model space. However, similar inconsistency phenomena also exist in data space. For instance, in ambient noise tomography, inter-receiver travel times are estimated from interferometric recordings by either averaging positive and negative noise cross-correlations or considering only the positive correlations. Such approaches introduce a condition that travel time from receiver A to receiver B equals to that from B to A in the data space. How this affects Bayesian travel time tomographic solutions should be investigated in future.

In the impedance inversion example and travel time tomography example, we have demonstrated that MAP solutions change under changes of parametrisations. In deterministic inverse problems, the optimal solution that minimises a regularised data misfit function is often found by linear(ised) methods. Provided that the algorithm finds the globally optimal solution, the deterministic inversion is equivalent to searching for the MAP solution within the parameter space \citep{tarantola2005inverse}. Since prior information in probabilistic inverse problems is often formulated as regularisation terms in deterministic inverse problems, and since the change of variable rule controlled by a Jacobian transformation is valid in both probabilistic and deterministic situations, the BK-inconsistency also exists in solutions to deterministic inverse problems under different parametrisations. That is to say, all equations and analyses discussed in this paper are pertinent to deterministic inversion. This can partly explain the deviations between inverse problem solutions found by different research groups using different parametrisations \citep{montelli2006catalogue, houser2008shear, simmons2010gypsum, obayashi2013finite, koelemeijer2016sp12rts, lu2019tx2019slab, van2020accelerated, hosseini2020global, ritsema2020heterogeneity, cui2024glad, thrastarson2024reveal}.

To summarise, the principle of invariance states that our choice of mathematics to represent a physical system does not change the system itself. Therefore, physically-consistent mathematical methods should all produce the same physical result. Perhaps our most important conclusions are therefore some key questions and problems that have emerged from this work:
\begin{itemize}[leftmargin=0.65cm]
	\item Is there a unique, natural parametrisation for any specific physical inference problem? Our work implies that the justification for such an assertion must be placed under intense scrutiny.
	\item By definition, maximum \textit{a posteriori} or MAP solutions provide the best fits to observed data and prior information. Yet, MAP solutions vary with the mathematical parametrisation, particularly for poorly constrained parameters which are most in need of prior constraints.
	\item \cite{mosegaard2024inconsistency} proved that common trans-dimensional inference algorithms produce contradictory results under simple changes in parametrisation. Specifically in tomographic problems, our work demonstrates that changes in grid cell sizes implied by changes in parametrisation lead directly to non-physical (inconsistent) results.
\end{itemize}

In statistical literature the BK-inconsistency is known, and typically said to be `resolvable'. This is because Bayes estimators are in fact unique under reparametrisation if, for each parametrisation, we choose an appropriate, so-called, \textit{sub-$\sigma$-algebra}. Loosely, this is a collection of sets which contain the lower-dimensional manifold, and which have a well-defined probability measure. Unfortunately, this does not solve the problem from a physical perspective: the choice of sub-$\sigma$-algebra is itself non-unique, and in general, different choices give different conditional probability densities \citep{gyenis2017conditioning}. Hence again, the calculated physical properties depend on the analyst's choice of mathematics, violating the principle of physical invariance.

The issues raised above are therefore serious and fundamental. The mathematical and statistical solutions to the BK-inconsistency do not hold water when considered in the context of physical science. The effects that we demonstrate herein are likely to affect past and present inverse problem solutions in a variety of different fields of application.

\section{Conclusion}
In this work, we have demonstrated the Borel-Kolmogorov inconsistency (BK-inconsistency) problem, in which conditional probability densities of the same information within parameter subspaces may be inconsistent, when evaluated under different parametrisations. This is because infinitely many ways exist to define probabilities over the same lower-dimensional parameter subspace manifold given those in a high-dimensional space, each leading to distinct conditional probability density distributions. Therefore, directly assigning values from high-dimensional distributions to lower-dimensional ones is only one choice out of infinitely many ways to define conditional probabilities. We show that the BK-inconsistency results in inconsistent conditional prior probability densities despite identical prior information being considered. We investigate their impact on Bayesian posterior solutions in seismic impedance inversion, surface wave dispersion inversion, and travel time tomography problems. Entirely different inversion results are obtained due to the BK-inconsistency, and it is difficult, if not impossible, to discriminate different such solutions from observed data alone. From the travel time tomography example, we find that reduction in data information tend to amplify these inconsistencies. Looking forward, it is crucial to investigate the extent to which the BK-inconsistency influences other commonly used Bayesian methods in geoscientific inference problems. We therefore suggest that careful consideration must be given when solving Bayesian inference problems that may involve changes in parametrisations.

\begin{acknowledgments}
XZ and AC thank the Edinburgh Imaging Project (EIP - \url{https://blogs.ed.ac.uk/imaging/}) sponsors (BP and TotalEnergies) for supporting this research. For the purpose of open access, we have applied a Creative Commons Attribution (CC BY) licence to any Author Accepted Manuscript version arising from this submission.
\end{acknowledgments}

\bibliographystyle{gji}
\bibliography{reference}

\label{lastpage}
\end{document}